\documentclass[10pt,reqno]{amsart}

\usepackage[
  a4paper,
  top=2.5cm,
  bottom=2.5cm,
  left=2.5cm,
  right=2.5cm
]{geometry}

\usepackage{amsmath}
\usepackage{amssymb}
\usepackage{amsthm}
\usepackage{mathtools}

\usepackage{graphicx}
\usepackage{booktabs}
\usepackage{array}
\usepackage{multirow}

\usepackage{indentfirst}
\usepackage{csquotes}
\usepackage{xcolor}
\usepackage{paralist}
\usepackage{etoolbox}
\usepackage{microtype}
\usepackage{titlesec}

\usepackage[numbers,sort&compress]{natbib}

\usepackage[
  colorlinks=true,
  linkcolor=black,
  citecolor=black,
  urlcolor=blue,
  filecolor=black
]{hyperref}

\usepackage{bookmark}
\usepackage{orcidlink}
\hypersetup{
  pdftitle={
    Dynamical and Observational Analysis of
    Generalized Nash's Theory of Gravity
  },
  pdfauthor={
    Amin Rezaei Akbarieh,
    Mohammad Amin Bolouri,
    Yaghoub Heydarzade
  },
  pdfsubject={Generalized Nash-Type Gravity},
  pdfkeywords={
    modified gravity,
    f(R,chi) gravity,
    dynamical systems,
    dark energy,
    cosmological constraints
  }
}

\titleformat{\section}[block]
  {\normalfont\Large\bfseries}
  {\thesection.}
  {0.75em}
  {}

\titleformat{\subsection}[block]
  {\normalfont\large\bfseries}
  {\thesubsection.}
  {0.75em}
  {}

\titleformat{\subsubsection}[block]
  {\normalfont\normalsize\bfseries}
  {\thesubsubsection.}
  {0.75em}
  {}

\titlespacing*{\section}
  {0pt}
  {3.5ex plus 1ex minus 0.2ex}
  {1.5ex plus 0.2ex}

\titlespacing*{\subsection}
  {0pt}
  {3ex plus 0.8ex minus 0.2ex}
  {1.2ex plus 0.2ex}

\titlespacing*{\subsubsection}
  {0pt}
  {2.5ex plus 0.6ex minus 0.2ex}
  {1ex plus 0.2ex}

\makeatletter
\renewcommand{\@setauthors}{%
  \begingroup
  \centering
  \normalfont\normalsize
  \authors\par
  \endgroup
  \vskip 1em
}
\makeatother

\begin{document}

% -------------------------------------------------
% Title
% -------------------------------------------------
\title[Generalized Nash Gravity]
{Dynamical and Observational Analysis of Generalized Nash's Theory of Gravity}

% -------------------------------------------------
% Authors and affiliations
% -------------------------------------------------
\author[A. Rezaei Akbarieh et al.]{%
Amin Rezaei Akbarieh\,
\orcidlink{0000-0002-4638-3751}$^{1,*}$
\quad
Mohammad Amin Bolouri\,
\orcidlink{0009-0002-1813-5088}$^{2,\dagger}$
\quad
Yaghoub Heydarzade\,
\orcidlink{0000-0001-7014-653X}$^{3,\ddagger}$
\\[0.7em]
{\small
$^{1}$Department of Physics,
Kocaeli University,
41001 Izmit,
T\"urkiye}
\\
{\small
$^{2}$Faculty of Physics,
University of Tabriz,
Tabriz,
Iran}
\\
{\small
$^{3}$Department of Mathematics,
Faculty of Sciences,
Bilkent University,
06800 Ankara,
T\"urkiye}
}

\date{}

% -------------------------------------------------
% Abstract
% -------------------------------------------------
\begin{abstract}
We investigate cosmic evolution in generalized Nash's theory of gravity involving the
quadratic Ricci invariant $\chi=R_{\mu\nu}R^{\mu\nu}$. The analysis is divided
into two complementary branches. First, we study the power-law family
$f(R,\chi)=R^{\alpha}+\beta\chi$ as a reduced autonomous system in a flat FLRW
background. Because the adopted variables become singular at the
Einstein--Hilbert limit $\alpha=1$, the phase-space analysis is restricted to
$\alpha\neq1$, with $\alpha=2$ used as a representative quadratic benchmark.
This benchmark contains radiation-like boundary configurations, restricted
scaling saddles, and de Sitter-like accelerating endpoints (a stable node away
from $\alpha=2$ and non-hyperbolic at the benchmark itself), but not a complete
regular radiation-to-matter-to-de Sitter sequence. Second, we constrain the
regular observational branch
$f_{\rm obs}(R,\chi)=R-2\Lambda+\beta\chi$, which reduces exactly to flat
$\Lambda$CDM when $\beta\to0$. The Hubble rate is obtained from the reduced
$\Lambda$CDM-connected background branch, integrated over $0\le z\le10$ and
matched at higher redshift to a standard radiation+matter+$\Lambda$ background.
Using SNe~Ia, BAO, and Planck~2018 compressed CMB distance priors, we find an
expansion history very close to $\Lambda$CDM, with the quadratic correction
tightly constrained around the nested standard-model limit. The resulting bound
on $\beta$ should be interpreted as a background-level constraint within this
reduced prescription, not as a perturbation-level viability test of the full
higher-derivative theory.
\end{abstract}

\maketitle

% -------------------------------------------------
% Author email footnotes
% -------------------------------------------------
\begingroup
\renewcommand{\thefootnote}{\fnsymbol{footnote}}

\footnotetext[1]{%
Corresponding author:
\href{mailto:amin.rezaeiakbarieh@kocaeli.edu.tr}
{\texttt{amin.rezaeiakbarieh@kocaeli.edu.tr}}}

\footnotetext[2]{%
Email:
\href{mailto:amingolami79@gmail.com}
{\texttt{amingolami79@gmail.com}}}

\footnotetext[3]{%
Email:
\href{mailto:yheydarzade@bilkent.edu.tr}
{\texttt{yheydarzade@bilkent.edu.tr}}}

\endgroup

% -------------------------------------------------
% Keywords below the abstract
% -------------------------------------------------
\vspace{-0.5em}

\begin{center}
\begin{minipage}{0.86\textwidth}
\end{minipage}
\end{center}

\vspace{0.5em}

% The main text continues from here

\section{Introduction}

In recent decades, modified theories of gravity have been extensively studied as
possible explanations for primordial inflation and the late-time acceleration of
the universe without introducing an additional dark-energy component
\citep{Clifton:2011jh,Nojiri:2017ncd}. Among the simplest extensions are
$f(R)$ theories \citep{Sotiriou:2008rp}, in which the Einstein--Hilbert
Lagrangian is replaced by a function of the Ricci scalar. A prominent example is
the Starobinsky model, $f(R)=R+R^2/(6M^2)$, which provides one of the earliest
and most successful geometric models of inflation \citep{Starobinsky:1980te}.
Scalar-tensor theories, such as the Brans--Dicke model, extend the gravitational
sector by introducing an additional scalar degree of freedom nonminimally coupled
to curvature \citep{Brans:1961sx}; in fact, metric $f(R)$ theories can be
recast as a particular class of scalar-tensor models. Observational evidence from
Type Ia supernovae
\citep{SupernovaCosmologyProject:1998vns,SupernovaSearchTeam:1998fmf,Pan-STARRS1:2017jku},
baryon acoustic oscillations \citep{SDSS:2005xqv,SDSS:2009ocz}, and the cosmic
microwave background \citep{Planck:2018vyg,Caldwell:2003hz} indicates that the
universe has undergone at least two accelerated phases: an early inflationary
epoch and the present dark-energy-dominated expansion. These observations impose
strong constraints on any viable modification of the gravitational action.

John Nash proposed a gravitational theory in an unpublished work, later cited in
\citep{Channuie:2018din}, whose Lagrangian density contains the quadratic
curvature combination $2R_{\mu\nu}R^{\mu\nu}-R^2$. Because the original action
does not contain the Einstein--Hilbert term, it does not possess the standard
weak-field GR limit in a straightforward way. To overcome this limitation,
extended versions have been proposed. For example, Channuie et al. considered a
model in which an Einstein--Hilbert term is added to the Nash action, allowing
the theory to recover Einstein gravity in weak-curvature regimes
\citep{Channuie:2018kfm}. Motivated by these developments, we consider
generalized Nash-type models involving both the Ricci scalar $R$ and the
quadratic Ricci invariant
$\chi=R_{\mu\nu}R^{\mu\nu}$. The theoretical phase-space analysis is performed
for the power-law family $f(R,\chi)=R^{\alpha}+\beta\chi$, while the
observational constraints are obtained for the regular Einstein--Hilbert branch
$R-2\Lambda+\beta\chi$.

A powerful approach to studying the qualitative evolution of cosmological models
is the dynamical-systems method, which rewrites the cosmological field equations
as an autonomous system of first-order differential equations
\citep{Copeland:2006wr,Bahamonde:2017ize,Boehmer:2014vea,Ribeiro:2014sla}.
Critical points of the autonomous system correspond to possible asymptotic or
transient cosmological regimes, such as radiation-like, matter-like, scaling, or
accelerated states. Their stability is determined by the eigenvalues of the
Jacobian matrix evaluated at each point: stable points describe attractors,
unstable points describe repellers, and saddle points correspond to transient
epochs. This method allows one to identify the global structure of the cosmic
phase space without solving the full set of background equations explicitly. In
Section~\ref{sec:dynamical}, we apply this approach to the power-law branch
$f(R,\chi)=R^{\alpha}+\beta\chi$ and analyse the corresponding critical points,
their physical admissibility, and their stability properties.

A complementary route is to confront the theory with background cosmological
data. Type Ia supernovae in the Pantheon+ compilation act as standard candles and
measure the luminosity-distance--redshift relation
\citep{Scolnic:2021amr,Brout:2022vxf}. BAO observations provide a standard ruler
that constrains transverse and radial distances at intermediate redshifts
\citep{eBOSS:2020yzd}. The Planck~2018 CMB distance priors summarize the
high-redshift geometric information encoded in the acoustic peaks
\citep{Planck:2018vyg}. Together, SNe~Ia, BAO, and CMB distance information place
tight constraints on the background expansion history and on possible deviations
from the standard $\Lambda$CDM model.

The observational analysis in this work is deliberately separated from the
autonomous-system construction. The latter is carried out for the power-law
family $f(R,\chi)=R^{\alpha}+\beta\chi$ in the regular sector $\alpha\neq1$, with
$\alpha=2$ used as a representative quadratic benchmark. By contrast, the
background likelihood analysis focuses on the regular Einstein--Hilbert branch
supplemented by a cosmological constant and a Ricci-tensor-squared correction,
$f_{\rm obs}(R,\chi)=R-2\Lambda+\beta\chi .$
We emphasize that these two parts address complementary questions within the same
generalized Nash-type framework, rather than two identical parameter slices. The
power-law family is useful for exposing the critical-point structure and the role
of the exponent $\alpha$ in generating accelerated endpoints. The observational
branch, on the other hand, is selected because it has a regular
Einstein--Hilbert limit and a nested flat-$\Lambda$CDM background when
$\beta\to0$.

This separation is essential for two reasons. First, the autonomous variables
used in Section~\ref{sec:dynamical} become singular at $\alpha=1$, while the
background Friedmann equations themselves remain regular for $f_R=1$. Second, the
term $-2\Lambda$ provides the standard late-time acceleration and guarantees that
the observational branch reduces exactly to $\Lambda$CDM when $\beta\to0$.
Therefore, the phase-space analysis should not be interpreted as a direct
stability proof of the fitted background branch. Instead, it provides theoretical
motivation for studying Ricci-tensor-squared corrections, while the data analysis
independently constrains the regular low-curvature branch. In this branch, the
parameter $\beta$ can be interpreted observationally as a direct measure of the
departure from the $\Lambda$CDM background induced by the invariant
$\chi=R_{\mu\nu}R^{\mu\nu}$.

In this setting, the dimensionless Hubble function $E(z)=H(z)/H_0$ is obtained
from the reduced background equation associated with the modified Friedmann
system, rather than from a purely phenomenological parameterization. We constrain
the matter density $\Omega_m$, the present Hubble constant $H_0$, the physical
baryon density $\omega_b h^2$, and the curvature-correction parameter $\beta$
through Bayesian inference with Markov chain Monte Carlo methods. The results,
presented in Sections~\ref{sec:obs} and \ref{sec:results}, quantify how tightly
current background data allow the Nash-type $\chi$ correction to deviate from the
nested $\Lambda$CDM limit.

This paper is structured as follows. In Section~\ref{sec:frx}, we introduce the
$f(R,\chi)$ theory and derive the modified Friedmann equations from its action.
In Section~\ref{sec:dynamical}, the corresponding autonomous dynamical system is
constructed for the power-law family $R^{\alpha}+\beta\chi$; its critical points
are determined, and their stability and physical admissibility are analysed.
In Section~\ref{sec:obs}, we present the observational framework for the regular
branch $R-2\Lambda+\beta\chi$, including the reduced background equation, the
numerical solution strategy, and the SNe~Ia, BAO, and CMB distance-prior
likelihoods. Section~\ref{sec:results} reports the resulting parameter
constraints, goodness-of-fit and model-selection diagnostics, and reconstructed
background evolution. Finally, Section~\ref{sec:conclusions} summarizes our
findings and discusses the limitations and implications of the analysis.

\section[The f(R,chi) Gravity]{The $f(R,\chi)$ Gravity}\label{sec:frx}

In this section, we introduce the $f(R, \chi)$ gravity and derive the corresponding field equations. The action of this gravitational theory is \citep{Feng:2019ejb}
\begin{equation}\label{eq:action}
S = \frac{1}{2\kappa^2} \int d^4x \sqrt{-g}  f(R, \chi) + \int d^4x\sqrt{-g} \mathcal{L}_m,
\end{equation}
where $f(R, \chi)$ is an arbitrary function of the Ricci scalar $R$ and the quadratic invariant term $\chi = R_{\mu\nu}R^{\mu\nu}$, $\mathcal{L}_m$ denotes the Lagrangian of matter and the constant $\kappa^2 = 8\pi G$ is defined in the system of natural units. By varying the action \eqref{eq:action} with respect to the metric $g_{\mu\nu}$, the modified gravitational field equations are obtained as
\begin{equation}\label{eq:field_eqn}
G_{\mu\nu} \equiv R_{\mu\nu} - \frac{1}{2} R g_{\mu\nu} = \frac{\kappa^2}{f_R} T_{\mu\nu} + \Pi_{\mu\nu},
\end{equation}
where $f_R \equiv \partial f/\partial R$ and $T_{\mu\nu}$ is the energy-momentum tensor of the matter fields. The symmetric tensor $\Pi_{\mu\nu}$ representing the additional high order corrections has the form

\begin{align}\label{eq:Omega}
\Pi_{\mu\nu} = {} & \frac{1}{f_R} \Bigl[
\frac{1}{2} g_{\mu\nu} (f - R f_R) + \nabla_\mu \nabla_\nu f_R \nonumber\\
& - g_{\mu\nu} \Box f_R
- 2 f_\chi R^\alpha_{\;\mu} R_{\alpha\nu} - \Box (f_\chi R_{\mu\nu}) \nonumber \\
& - g_{\mu\nu} \nabla_\alpha \nabla_\beta (f_\chi R^{\alpha\beta}) \nonumber\\
& + 2 \nabla_\alpha \nabla_{(\mu} (R^\alpha_{\;\nu)} f_\chi)
\Bigr],
\end{align}
where $f_\chi \equiv \partial f/\partial \chi$, and $\Box \equiv g^{\alpha\beta} \nabla_\alpha \nabla_\beta$ denotes the d'Alembert operator. For a homogeneous and isotropic universe, we consider the flat FLRW metric

\begin{equation}\label{eq:metric}
ds^2 = -dt^2 + a^2(t) \left( dr^2 + r^2 d\Omega^2 \right),
\end{equation}
where $a(t)$ is the scale factor in which the Ricci scalar and the quadratic invariant $\chi$ in terms of the Hubble parameter $H = \dot{a}/a$ become

\begin{equation}\label{eq:Rchi}
R = 6 (\dot{H} + 2 H^2), \quad \chi = 12 \left( 3H^4 + 3 H^2 \dot{H} + \dot{H}^2 \right),
\end{equation}
where dots denote derivatives with respect to the cosmic time $t$.

\paragraph{Dimensional consistency.}
Since $R\sim L^{-2}$ while $\chi=R_{\mu\nu}R^{\mu\nu}\sim L^{-4}$, the two terms
in $f(R,\chi)=R^{\alpha}+\beta\chi$ carry different curvature dimensions unless a
reference scale is introduced. To make the action dimensionally well defined we
introduce a constant reference curvature scale $R_\star$ (of order $H_0^2$ for the
late-time application of interest) and write the model in the manifestly
consistent form
\begin{equation}\label{eq:dimensionless_f}
f(R,\chi)=R_\star\left[\left(\frac{R}{R_\star}\right)^{\alpha}
+\beta\,\frac{\chi}{R_\star^{2}}\right],
\end{equation}
in which both bracketed terms are dimensionless and $f$ has the correct curvature
dimension. The deviation parameter $\beta$ quoted throughout the observational
analysis is therefore a \emph{dimensionless} coefficient measured relative to the
scale $R_\star$. For compactness we continue to write the model as
$f(R,\chi)=R^{\alpha}+\beta\chi$, with the understanding that all curvature
quantities are expressed in units of $R_\star$.

The $00$-component of the field equations \eqref{eq:field_eqn} yields the modified Friedmann equation

\begin{align}
3 f_R H^2 &= \kappa^2 \rho - 3 H \dot{f_R} + \frac{1}{2} (R f_R - 2\chi f_{\chi}- f) \nonumber\\
&\quad -  \dot{f_\chi}(18 H^3 + 12 H \dot{H}) \nonumber \\
& \quad - 6 f_\chi \left( 3 H^4 - H^2 \dot{H} - 4 \dot{H}^2 + 2 H \ddot{H} \right), \label{eq:friedmann1}
\end{align}
while the spatial $ii$-components give
\begin{align}
f_R (3 H^2 + 2 \dot{H}) &= -\kappa^2 p - 4 H \dot{f_R} - \ddot{f_R} \nonumber\\
&\quad - \frac{1}{2} (R f_R - f) \nonumber \\
& \quad + R \ddot{f_\chi} + \dot{f_\chi} (2 H R + \dot{R}) \nonumber \\
& \quad + 6 f_\chi \bigl( -15 H^4 + 5 H^2 \dot{H} - 6 \dot{H}^2 \nonumber\\
&\qquad\qquad + 12 H \ddot{H} + 2 \dddot{H} \bigr),\label{eq:friedmann2}
\end{align}
where $\rho$ and $p$ represent the total energy density and pressure of the cosmic fluid, respectively.\\
From a physical point of view, the presence of higher-order curvature terms, in particular the invariant $\chi$ term, in the gravitational equations introduces new dynamic degrees of freedom into the gravitational part of the theory \citep{Clifton:2011jh,Stelle:1977ry}. Unlike general relativity (GR), where the dynamics of gravity depend only on the Ricci scalar $R$, in the $f(R,\chi)$ model the gravitational behavior depends more complexly on the local structure of the curvature of spacetime and can be sensitive to anisotropies or tidal forces \citep{Sotiriou:2008rp,Nojiri:2010wj}. These extra conditions can be viewed as significant extensions of gravity theory that gain increasing importance in regimes with high curvature, e.g., the early universe or near massive objects \citep{Kolb:1990vq,KAGRA:2021vkt}. Cosmologically, these modifications can have a dramatic influence on the history of the universe, offering the possibility of a geometric account of both early inflation and late time acceleration of the universe \citep{Baumann:2009ds,Caldwell:2009ix}. The altered Friedmann equations obtained in this section unify these impacts and serve as the theoretical basis for the dynamic analysis in the following sections of the paper.

\section{Dynamical System Analysis}\label{sec:dynamical}

Rather than solving the modified Friedmann equations case by case, we recast the
cosmological dynamics as an autonomous system and analyse the qualitative
structure of its phase space. In this language, radiation-like, scaling, and
accelerating regimes correspond to critical points of the flow. Their local
stability is determined by the eigenvalues of the Jacobian matrix evaluated at
each point: eigenvalues with negative real parts indicate an attractor,
eigenvalues with positive real parts indicate a repeller, and a mixture of signs
indicates a saddle. When one or more eigenvalues vanish, the point is
non-hyperbolic and linear stability analysis is not sufficient; a centre-manifold
analysis would then be required for a definitive classification.

We specialize the dynamical construction to the power-law Nash-type family
\begin{equation}\label{f}
f(R,\chi)=R^{\alpha}+\beta\chi,
\end{equation}
where $\chi=R_{\mu\nu}R^{\mu\nu}$. This branch is used to diagnose the qualitative
role of the Ricci-tensor-squared correction in the phase-space structure. It
should be distinguished from the observational branch
$R-2\Lambda+\beta\chi$, which is studied separately in
Section~\ref{sec:obs} because it has a regular Einstein--Hilbert limit and a
nested $\Lambda$CDM background.

\subsection{Dimensionless variables}\label{subsec:dyn_variables}

For the dynamical analysis we introduce
\begin{equation}\label{eq:vars}
\begin{aligned}
\Omega_r &= \frac{\kappa^2 \rho_r}{3 H^2 f_R}, &
\Omega_m &= \frac{\kappa^2 \rho_m}{3 H^2 f_R}, \\
x_1 &= \frac{\dot{f_R}}{H f_R}, &
x_2 &= \frac{R}{6 H^2}, \\
x_3 &= \frac{\beta \chi}{3 f_R H^2}, &
x_4 &= \frac{f}{6 H^2 f_R}, \\
x_5 &= \frac{6 \beta H^2}{f_R}, &
x_6 &= \frac{18 \beta \dot{H}}{f_R}, \\
x_7 &= \frac{8 \beta \dot{H}^2}{H^2 f_R}, &
x_8 &= \frac{2 \beta \dot{R}}{3H f_R} .
\end{aligned}
\end{equation}
The density parameters in Eq.~\eqref{eq:vars} are normalized by the effective
gravity coupling $f_R$ rather than by $3H^2$. Therefore $\Omega_m$ and
$\Omega_r$ are effective phase-space variables and should not automatically be
identified with the usual fractional densities of GR. This distinction is
important near boundaries where $f_R$ is singular or vanishes.

For the power-law branch one has
\begin{equation}
f_R=\alpha R^{\alpha-1}.
\end{equation}
The autonomous chart is therefore regular only where $f_R$ is finite and
nonzero. In practice we work with $\alpha\neq0,1$ and $R\neq0$ in the regular
interior of the phase space. For $\alpha>1$, the locus $R=0$, equivalently
$x_2=0$, is a singular boundary because $f_R\to0$; for $\alpha<1$ the same locus
is an asymptotic boundary because $f_R$ diverges. Fixed points located at
$x_2=0$ must therefore be interpreted with this chart-dependent caveat.

Substitution of Eq.~\eqref{eq:vars} into the modified Friedmann equation
\eqref{eq:friedmann1} gives the constraint
\begin{equation}\label{eq:constraint}
\Omega_m+\Omega_r-x_1+x_2-x_3-x_4-x_5+x_6+x_7-x_8=1 .
\end{equation}
Using $N=\ln a$ as the time variable, the evolution equations are
\begin{align}
\Omega'_m &= \Omega_m\left(-3-x_1-2\frac{\dot H}{H^2}\right),\label{eq:omdot}\\
\Omega'_r &= \Omega_r\left(-4-x_1-2\frac{\dot H}{H^2}\right),\label{eq:ordot}\\
x'_1 &= \frac{\ddot f_R}{H^2 f_R}-x_1^2-x_1\frac{\dot H}{H^2},\label{eq:x1dot}\\
x'_2 &= \frac{\dot R}{6H^3}-2x_2\frac{\dot H}{H^2},\label{eq:x2dot}\\
x'_3 &= x_3\left(\frac{\dot\chi}{\chi H}-x_1-2\frac{\dot H}{H^2}\right),\label{eq:x3dot}\\
x'_4 &= \frac{\dot f}{6f_RH^3}-x_1x_4-2x_4\frac{\dot H}{H^2},\label{eq:x4dot}\\
x'_5 &= x_5\left(2\frac{\dot H}{H^2}-x_1\right),\label{eq:x5dot}\\
x'_6 &= x_6\left(\frac{\ddot H}{H\dot H}-x_1\right),\label{eq:x6dot}\\
x'_7 &= x_7\left(2\frac{\ddot H}{H\dot H}-2\frac{\dot H}{H^2}-x_1\right),\label{eq:x7dot}\\
x'_8 &= x_8\left(\frac{\ddot R}{H\dot R}-\frac{\dot f_R}{Hf_R}-\frac{\dot H}{H^2}\right).\label{eq:x8dot}
\end{align}
From the FLRW identities in Eq.~\eqref{eq:Rchi}, the auxiliary ratios entering
these equations can be expressed in terms of the phase-space variables as
\begin{equation}\label{eq:HdotH2}
\frac{\dot H}{H^2}=x_2-2,
\end{equation}
\begin{align}
\frac{\dot R}{6H^3} &= \frac{x_1x_2}{\alpha-1},\label{eq:Rdot}\\
\frac{\dot f}{6f_RH^3} &=
\frac{1}{3}\left[-2(x_2-2)^2x_5+
\frac{x_1x_2\bigl(3+x_5(2x_2-1)\bigr)}{\alpha-1}\right],\label{eq:fdot}\\
\frac{\dot\chi}{\chi H} &=
\frac{8+x_1x_2(2x_2-1)-2(x_2-4)x_2(\alpha-1)-8\alpha}
{\bigl[1+(x_2-1)x_2\bigr](\alpha-1)} .\label{eq:chidot}
\end{align}
The factor $1+(x_2-1)x_2=x_2^2-x_2+1$ is strictly positive for all real $x_2$ and
therefore introduces no additional real singularity. The singularity at
$\alpha=1$ is instead a property of the chosen autonomous variables: for
$\alpha=1$, $f_R$ is constant, $x_1$ vanishes identically, and
Eqs.~\eqref{eq:Rdot}--\eqref{eq:chidot} take indeterminate $0/0$ forms. The
underlying Friedmann equations remain regular in the Einstein--Hilbert branch;
this is why the observational analysis is performed separately in
Section~\ref{sec:obs}.

The variables $x_1,x_3,x_4,x_6,x_7$, and $x_8$ are algebraically determined by
$\{\Omega_m,\Omega_r,x_2,x_5\}$. Explicitly,
\begin{align}
x_3 &= \frac{2}{3}x_5(x_2^2-x_2+1),\label{eq:x3def}\\
x_4 &= \frac{x_2}{\alpha}+\frac{1}{3}x_5(x_2^2-x_2+1),\label{eq:x4def}\\
x_6 &= 3x_5(x_2-2),\label{eq:x6def}\\
x_7 &= \frac{4}{3}x_5(x_2-2)^2,\label{eq:x7def}\\
x_8 &= \frac{2}{3(\alpha-1)}x_1x_2x_5 .\label{eq:x8def}
\end{align}
Inserting these relations into Eq.~\eqref{eq:constraint} yields
{\scriptsize
\begin{equation}\label{eq:x1def}
x_1=
\frac{(\alpha-1)\left[3\alpha(\Omega_m+\Omega_r+x_2-1)-3x_2
+\alpha x_5(x_2^2-4x_2-8)\right]}
{\alpha(3\alpha+2x_2x_5-3)} .
\end{equation}
}
The hypersurface on which the denominator of Eq.~\eqref{eq:x1def} vanishes is
not part of this reduced chart. Away from this hypersurface, the four equations
for $\{\Omega_m,\Omega_r,x_2,x_5\}$ form a closed autonomous system, and the
remaining variables are reconstructed algebraically.

The kinematic observables are especially simple in terms of $x_2$. Since
\begin{equation}
x_2=\frac{R}{6H^2}=2+\frac{\dot H}{H^2},
\end{equation}
one obtains
\begin{equation}\label{eq:qweff}
q=-1-\frac{\dot H}{H^2}=1-x_2,
\qquad
w_{\rm eff}=\frac{2q-1}{3}=\frac{1-2x_2}{3} .
\end{equation}
Accelerated expansion requires $x_2>1$, and an exact de Sitter state corresponds
to $x_2=2$, for which $q=-1$ and $w_{\rm eff}=-1$.

\subsection{Critical points and stability}\label{sec:fixedpoints}

The critical points are obtained from
$\Omega'_m=\Omega'_r=x'_2=x'_5=0$. Table~\ref{table:fixpoints-existence} lists the
eight equilibria of the reduced system. Here ``existence'' means that the point
is a real solution of the autonomous equations, while ``admissibility'' also
requires non-negative effective density variables and membership in the regular
chart wherever applicable.

\begin{table*}[ht]
\centering
\small
\renewcommand{\arraystretch}{1.25}
\resizebox{\textwidth}{!}{%
\begin{tabular}{lccccccc}
\toprule
\textbf{Point} & $\Omega_m$ & $\Omega_r$ & $x_2$ & $x_5$ & $q$ & $w_{\rm eff}$ & \textbf{Existence / admissibility} \\
\midrule
$P_1$ & $0$ & $1$ & $0$ & $0$ & $1$ & $1/3$ & all $\alpha\neq0,1$; boundary for $\alpha>1$ \\
$P_2$ & $2$ & $0$ & $0$ & $0$ & $1$ & $1/3$ & all $\alpha\neq0,1$; boundary for $\alpha>1$ \\
$P_3$ & $0$ & $0$ & $0$ & $9/8$ & $1$ & $1/3$ & all $\alpha\neq0,1$; boundary for $\alpha>1$ \\
$P_4$ & $0$ & $0$ & $2$ & $(\alpha-2)/(4\alpha)$ & $-1$ & $-1$ & $\alpha\notin\left\{0,1,\frac{1\pm\sqrt{7}}{3}\right\}$ \\
$P_5$ & $0$ & $\dfrac{-2+8\alpha-5\alpha^2}{\alpha^2}$ & $2-\dfrac{2}{\alpha}$ & $0$ & $\dfrac{2-\alpha}{\alpha}$ & $\dfrac{4-3\alpha}{3\alpha}$ & $\alpha\in\left[\dfrac{4-\sqrt6}{5},\dfrac{4+\sqrt6}{5}\right]$, $\alpha\neq1$ \\
$P_6$ & $\dfrac{-3+(13-8\alpha)\alpha}{2\alpha^2}$ & $0$ & $2-\dfrac{3}{2\alpha}$ & $0$ & $\dfrac{3-2\alpha}{2\alpha}$ & $\dfrac{1-\alpha}{\alpha}$ & $\alpha\in\left[\dfrac{13-\sqrt{73}}{16},\dfrac{13+\sqrt{73}}{16}\right]$, $\alpha\neq1$ \\
$P_7$ & $0$ & $0$ & $\dfrac{\alpha(4\alpha-5)}{2\alpha^2-3\alpha+1}$ & $0$ & $1-\dfrac{\alpha(4\alpha-5)}{2\alpha^2-3\alpha+1}$ & $\dfrac{1}{3}\left[1-\dfrac{2\alpha(4\alpha-5)}{2\alpha^2-3\alpha+1}\right]$ & $\alpha\neq0,\frac12,1$ \\
$P_8$ & $0$ & $0$ & $0$ & $0$ & $1$ & $1/3$ & all $\alpha\neq0,1$; boundary for $\alpha>1$ \\
\bottomrule
\end{tabular}%
}
\caption{Critical points of the reduced autonomous system. The variables
$\Omega_m$ and $\Omega_r$ are normalized by $f_R$ and are therefore effective
phase-space densities. The admissibility windows of $P_5$ and $P_6$ follow from
$\Omega_r\ge0$ and $\Omega_m\ge0$, respectively. The points
$P_1,P_2,P_3$, and $P_8$ lie at $x_2=0$, i.e.\ $R=0$; for $\alpha>1$ this is a
singular boundary of the chosen chart because $f_R=\alpha R^{\alpha-1}$ vanishes.
Their stability classification should therefore be read as limiting information
about the flow approaching that boundary.}
\label{table:fixpoints-existence}
\end{table*}

\begin{table*}[ht]
\centering
\small
\renewcommand{\arraystretch}{1.25}
\resizebox{\textwidth}{!}{%
\begin{tabular}{l c c c l}
\toprule
\textbf{Point} &
\textbf{Regular chart?} &
\textbf{Physical admissibility} &
\textbf{Stability role} &
\textbf{Interpretation} \\
\midrule
$P_1$ & singular for $\alpha>1$ & radiation density positive formally & saddle & radiation-like boundary state \\
$P_2$ & singular for $\alpha>1$ & effective matter unusual & range-dependent & radiation-like effective boundary state \\
$P_3$ & singular for $\alpha>1$ & curvature dominated & generically unstable & $\chi$-dominated boundary state \\
$P_4$ & regular for $R\neq0$ & vacuum/de Sitter & range-dependent (typically saddle); non-hyperbolic at $\alpha=2$ & de Sitter state, degenerate with $P_7$ at $\alpha=2$ \\
$P_5$ & regular if $R\neq0$ & restricted $\alpha$ window & saddle & radiation/scaling transition state \\
$P_6$ & regular if $R\neq0$ & restricted $\alpha$ window & saddle & matter/scaling transition state \\
$P_7$ & regular if $R\neq0$ & vacuum & stable node for $\alpha\neq2$; non-hyperbolic at $\alpha=2$ & stable de Sitter/accelerating endpoint \\
$P_8$ & singular for $\alpha>1$ & vacuum boundary & saddle-like & radiation-like boundary state \\
\bottomrule
\end{tabular}%
}
\caption{Physical interpretation of the fixed points. The ``regular chart``
column refers to the autonomous variables in Eq.~\eqref{eq:vars}. For the
illustrative benchmark $\alpha=2$, the points at $x_2=0$ are limiting boundary
configurations rather than regular interior equilibria.}
\label{table:fixpoints-interpretation}
\end{table*}

For each point, the stability is determined from the Jacobian of the
four-dimensional reduced system with variables
$y=\{\Omega_m,\Omega_r,x_2,x_5\}$. This reduced Jacobian is the relevant one for
the stability classification. If the algebraically constrained variables
$\{x_1,x_3,\ldots,x_8\}$ are treated as independent, spurious zero eigenvalues
appear because of the constraints in Eqs.~\eqref{eq:x3def}--\eqref{eq:x1def}.

\subsection{Phase-space interpretation}\label{subsec:phase_space}

\begin{figure}[htbp]
\centering
\includegraphics[width=0.95\linewidth]{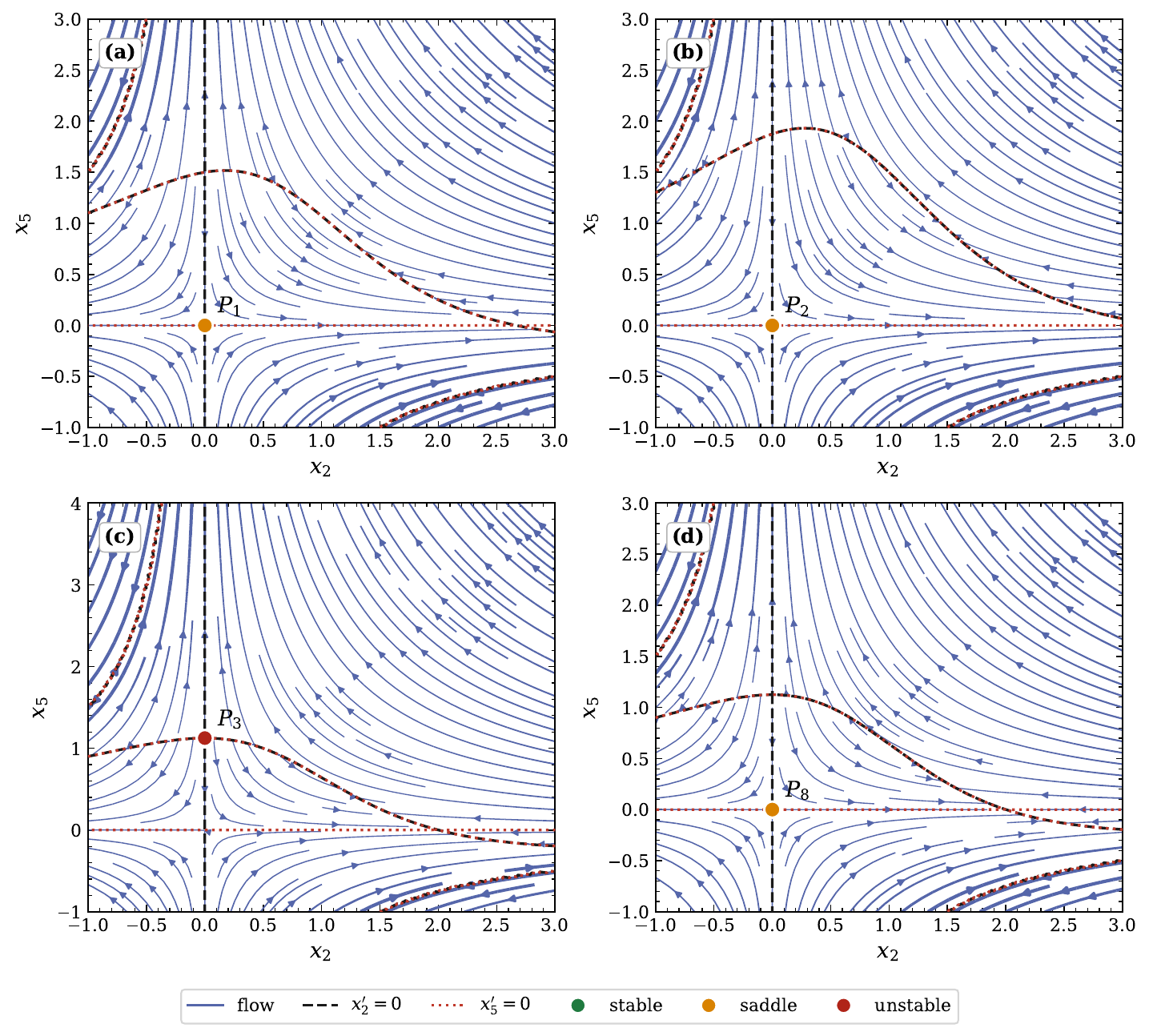}
\caption{Projected phase portraits in the $(x_2,x_5)$ plane for the early-time
sector $P_1$, $P_2$, $P_3$, and $P_8$. The markers indicate the fixed points,
dashed curves denote the projected nullclines, and streamlines show the local
flow. Since these points lie at $x_2=0$ ($R=0$), where $f_R$ vanishes for the
benchmark $\alpha=2$, the plots show the limiting flow geometry near the
singular boundary of the chart.}
\label{fig:early_phase}
\end{figure}

\begin{figure}[htbp]
\centering
\includegraphics[width=0.95\linewidth]{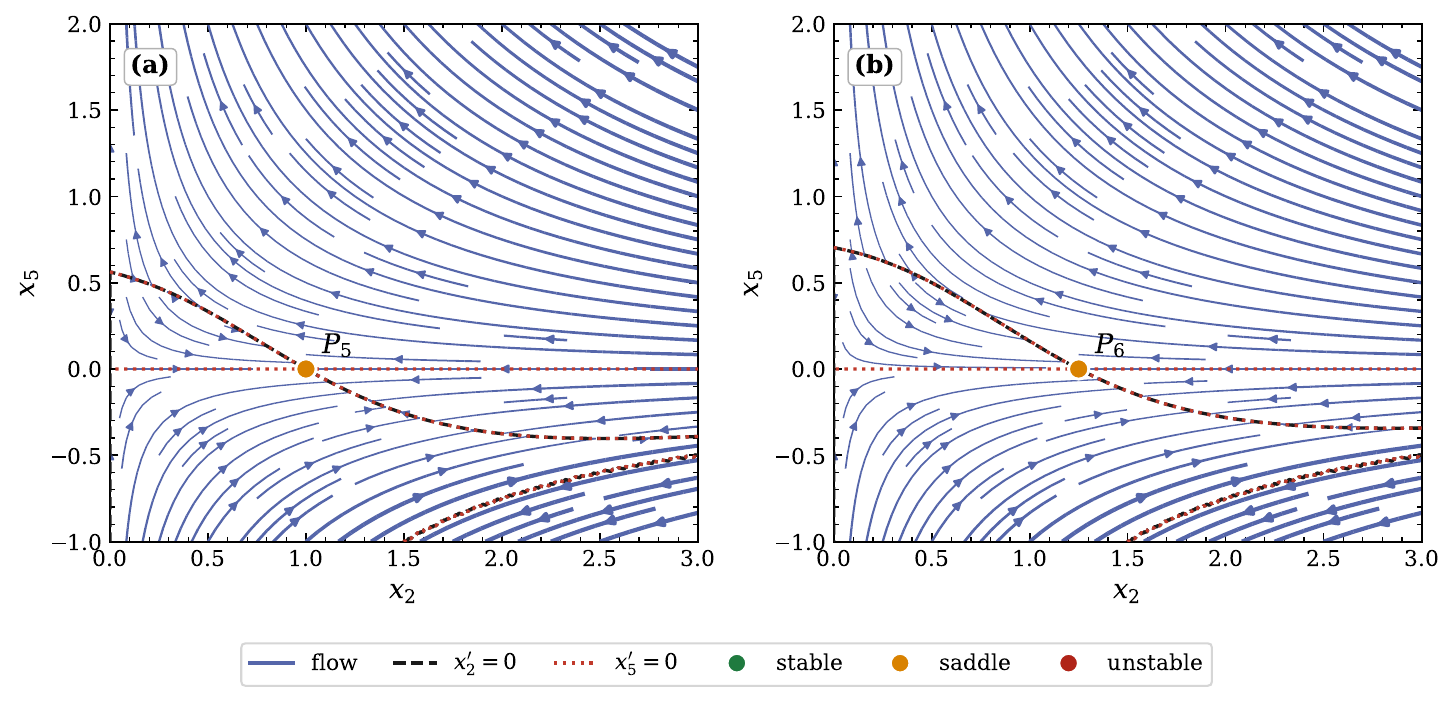}
\caption{Projected phase portraits in the $(x_2,x_5)$ plane for the intermediate
scaling sector $P_5$ and $P_6$. The trajectories pass through the neighbourhood
of these equilibria without converging to them, illustrating their role as local
transition states in the projected flow.}
\label{fig:trans_phase}
\end{figure}

\begin{figure}[htbp]
\centering
\includegraphics[width=0.95\linewidth]{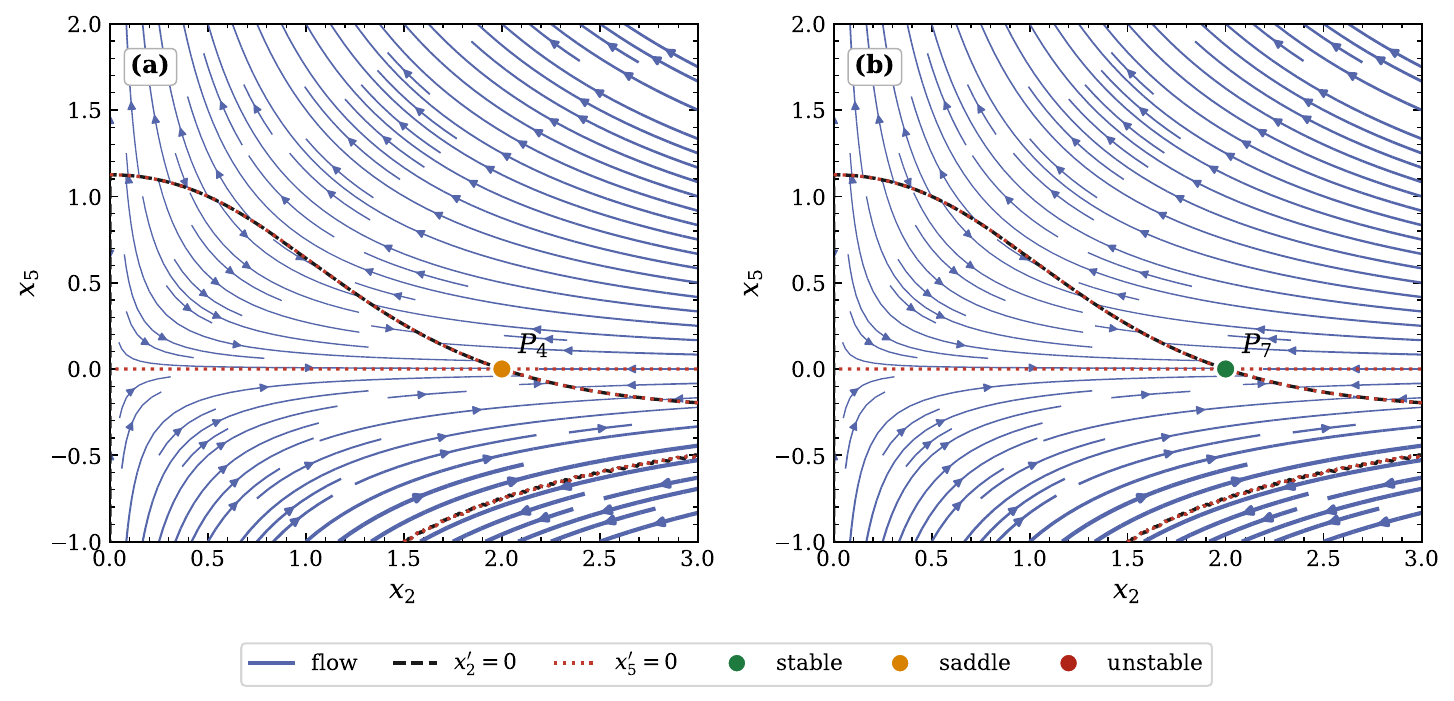}
\caption{Projected phase portraits in the $(x_2,x_5)$ plane for the accelerating
sector $P_4$ and $P_7$. At $\alpha=2$, the two branches share the same de Sitter
kinematics $x_2=2$ and become degenerate in the projected phase space.}
\label{fig:late_phase}
\end{figure}

The equilibria can be grouped into an early-time boundary sector
($P_1,P_2,P_3,P_8$), an intermediate scaling sector ($P_5,P_6$), and a late-time
accelerating sector ($P_4,P_7$). The phase portraits in
Figs.~\ref{fig:early_phase}--\ref{fig:late_phase} are two-dimensional projections
onto the $(x_2,x_5)$ plane, with $\Omega_m$ and $\Omega_r$ fixed at the
corresponding fixed-point values. They should therefore be interpreted as local
visualizations of the projected flow, not as global portraits of the full
four-dimensional phase space. The plots are shown for the representative
benchmark $\alpha=2$, while the analytic stability statements are given as
functions of $\alpha$.

\paragraph{Early-time boundary sector.}
The points $P_1,P_2,P_3$, and $P_8$ all have radiation-like kinematics,
$q=1$ and $w_{\rm eff}=1/3$, but they differ in effective matter and curvature
content. Since all four sit at $R=0$, they are boundary configurations of the
chosen chart for $\alpha>1$.

For $P_1=(0,1,0,0)$ the reduced-Jacobian eigenvalues are
\begin{equation}
\{1,-1,4,-4\},
\end{equation}
so $P_1$ is a saddle. It represents a radiation-like limiting configuration.

The point $P_2=(2,0,0,0)$ has the same radiation-like kinematics but a nonzero
effective matter variable. Because it lies on the $R=0$ boundary, the value
$\Omega_m=2$ should not be interpreted as an ordinary physical matter fraction.
Its eigenvalues are
\begin{equation}
\left\{-2,-1,-5,\frac{4\alpha-3}{\alpha-1}\right\},
\end{equation}
so its stability is range-dependent and it is not a robust matter attractor.

The point $P_3=(0,0,0,9/8)$ is dominated by the $\chi$ contribution. Its
eigenvalues are
\begin{equation}
\left\{5,4,3,\frac{4(\alpha-2)}{\alpha-1}\right\}.
\end{equation}
Because three eigenvalues are positive, $P_3$ is generically unstable and acts as
a curvature-dominated repelling boundary configuration.

Finally, $P_8=(0,0,0,0)$ has eigenvalues
\begin{equation}
\left\{2,1,-3,\frac{4\alpha-5}{\alpha-1}\right\},
\end{equation}
which are of mixed sign over broad parameter ranges. It is therefore a
saddle-like radiation boundary state. Collectively, the early-time sector should
be understood as a set of radiation-like limiting configurations, not as a fully
regular radiation epoch in the interior of the adopted variables.

\paragraph{Intermediate scaling sector.}
The points $P_5$ and $P_6$ represent scaling configurations that can organize
intermediate transitions only within restricted admissibility windows.
For $P_5$, one has
\begin{equation}
x_2=2-\frac{2}{\alpha},\qquad
q=\frac{2-\alpha}{\alpha},\qquad
w_{\rm eff}=\frac{4-3\alpha}{3\alpha} .
\end{equation}
Its eigenvalues are
\begin{equation}
\left\{1,\frac{4(\alpha-2)}{\alpha},
\frac{(\alpha-2)\pm\sqrt{3}\sqrt{27\alpha^2-44\alpha+12}}{2\alpha}\right\}.
\end{equation}
The presence of eigenvalues with different signs in the admissible range implies
that $P_5$ is a saddle. Its effective radiation density,
\begin{equation}
\Omega_r=\frac{-2+8\alpha-5\alpha^2}{\alpha^2},
\end{equation}
is non-negative only for
\begin{equation}
\alpha\in\left[\frac{4-\sqrt6}{5},\frac{4+\sqrt6}{5}\right],
\end{equation}
with $\alpha=1$ excluded by the chart. At the illustrative value $\alpha=2$,
$\Omega_r=-3/2$, so this point exists algebraically but is physically
inadmissible.

For $P_6$, one has
\begin{equation}
x_2=2-\frac{3}{2\alpha},\qquad
q=\frac{3-2\alpha}{2\alpha},\qquad
w_{\rm eff}=\frac{1-\alpha}{\alpha} .
\end{equation}
The eigenvalues are
{\footnotesize
\begin{equation}
\left\{-1,\frac{3(\alpha-2)}{\alpha},
\frac{-3\pm\dfrac{\sqrt{(\alpha-1)(256\alpha^3-608\alpha^2+417\alpha-81)}}{\alpha-1}}{4\alpha}\right\}.
\end{equation}
}
This point is also a saddle in the relevant parameter ranges. Its effective
matter density,
\begin{equation}
\Omega_m=\frac{-3+(13-8\alpha)\alpha}{2\alpha^2},
\end{equation}
is non-negative only for
\begin{equation}
\alpha\in\left[\frac{13-\sqrt{73}}{16},\frac{13+\sqrt{73}}{16}\right],
\end{equation}
again with $\alpha=1$ excluded. This window
($\alpha\in[0.28,1.35]$ numerically) excludes the benchmark $\alpha=2$, at which
$\Omega_m=-9/8<0$, so, like $P_5$, the point $P_6$ is physically inadmissible at
$\alpha=2$ and the portraits in Fig.~\ref{fig:trans_phase} show it only as a
transient projected configuration. Thus $P_5$ and $P_6$ are best interpreted as
scaling saddles that organize transitions only in restricted regions of the
parameter space.

\paragraph{Late-time accelerating sector.}
The late-time sector is governed by $P_4$ and $P_7$. The point $P_4$ has
$x_2=2$, and therefore
\begin{equation}
q=-1,
\qquad
w_{\rm eff}=-1 .
\end{equation}
It describes an exact de Sitter configuration. Since $x_2=2$ implies
$R=12H^2\neq0$, this point lies in the regular interior of the chart for
$\alpha\neq0,1$. Its eigenvalues are
{\scriptsize
\begin{equation}
\left\{-4,-3,
\frac{-(9\alpha^2-5\alpha-8)\pm
\sqrt{225\alpha^4-762\alpha^3+745\alpha^2+80\alpha-320}}
{2(3\alpha^2-2\alpha-2)}\right\}.
\end{equation}
}
It is also important to note that the existence of $P_4$ in this regular chart requires the denominator of $x_1$ in Eq.~\eqref{eq:x1def} to be non-zero. Substituting $x_2=2$ and $x_5=(\alpha-2)/(4\alpha)$ into the denominator $\alpha(3\alpha+2x_2x_5-3)$ yields $3\alpha^2-2\alpha-2$. Therefore, the values $\alpha=(1\pm\sqrt{7})/3$ must also be excluded from the domain of validity for $P_4$.
At $\alpha=2$, these become
\begin{equation}
\{-4,-3,-3,0\},
\end{equation}
where the vanishing eigenvalue arises because the two square-root roots collapse
to $\{0,-3\}$ at $\alpha=2$. The zero eigenvalue signals that $P_4$ is
non-hyperbolic at the benchmark, where it coincides with $P_7$ at
$(\Omega_m,\Omega_r,x_2,x_5)=(0,0,2,0)$ (see below). For most values of
$\alpha\neq2$, the non-trivial pair of roots contains at least one eigenvalue
with a positive real part, making $P_4$ a saddle. However, this is not
strictly true for all $\alpha$; in specific restricted intervals (e.g.,
$\alpha<0$), the real parts can be negative, meaning the stability of this
branch is generally range-dependent. The genuinely stable de Sitter endpoint
of the power-law sector is instead realized by the $P_7$ branch discussed
next.

The point $P_7$ is a second accelerating branch with
\begin{equation}
x_2=\frac{\alpha(4\alpha-5)}{2\alpha^2-3\alpha+1} .
\end{equation}
Its eigenvalues are
\begin{align}\label{eq:p7eig}
\Biggl\{\ &
-\frac{8\alpha^2-13\alpha+3}{(\alpha-1)(2\alpha-1)},\ \
-\frac{2(5\alpha^2-8\alpha+2)}{(\alpha-1)(2\alpha-1)}, \nonumber\\
&-\frac{4\alpha-5}{\alpha-1},\ \
-\frac{2(\alpha-2)^2}{(\alpha-1)(2\alpha-1)}
\Biggr\}.
\end{align}
At $\alpha=2$, this branch also has $x_2=2$, $q=-1$, and $w_{\rm eff}=-1$, so it
coincides with $P_4$ at $(\Omega_m,\Omega_r,x_2,x_5)=(0,0,2,0)$: the two
accelerating branches merge into a single de Sitter point. Precisely at
$\alpha=2$ the last eigenvalue vanishes and the shared point is non-hyperbolic,
with spectrum $\{-4,-3,-3,0\}$, so linear analysis alone cannot establish
asymptotic stability along the degenerate direction. Away from $\alpha=2$, the
branches separate: within the admissible range the four eigenvalues of $P_7$ in
Eq.~\eqref{eq:p7eig} are all negative, so $P_7$ is a \emph{stable node} and
constitutes the genuine late-time attractor of the power-law sector, while the
stability of $P_4$ remains range-dependent, being a saddle for most values of
$\alpha\neq2$ but not in specific restricted intervals such as $\alpha<0$.

The coordinate $x_5$ also requires care. By definition,
\begin{equation}
x_5=\frac{6\beta H^2}{f_R} .
\end{equation}
For $\alpha=2$, since $f_R=2R$ and $R=6H^2x_2$, this becomes
\begin{equation}
x_5=\frac{\beta}{2x_2} .
\end{equation}
Thus, at finite nonzero curvature, a fixed point with $x_5=0$ lies on the
$\beta=0$ invariant subspace of the power-law autonomous chart. Such points
should not be used to infer the observational value of $\beta$. They describe
limiting configurations of the power-law phase space in which the explicit
$\chi$ contribution is absent or dynamically suppressed. The observational
constraint on a nonzero $\beta$ is obtained independently from the regular
Einstein--Hilbert branch studied in Section~\ref{sec:obs}.

\begin{figure}[htbp]
\centering
\includegraphics[width=0.95\linewidth]{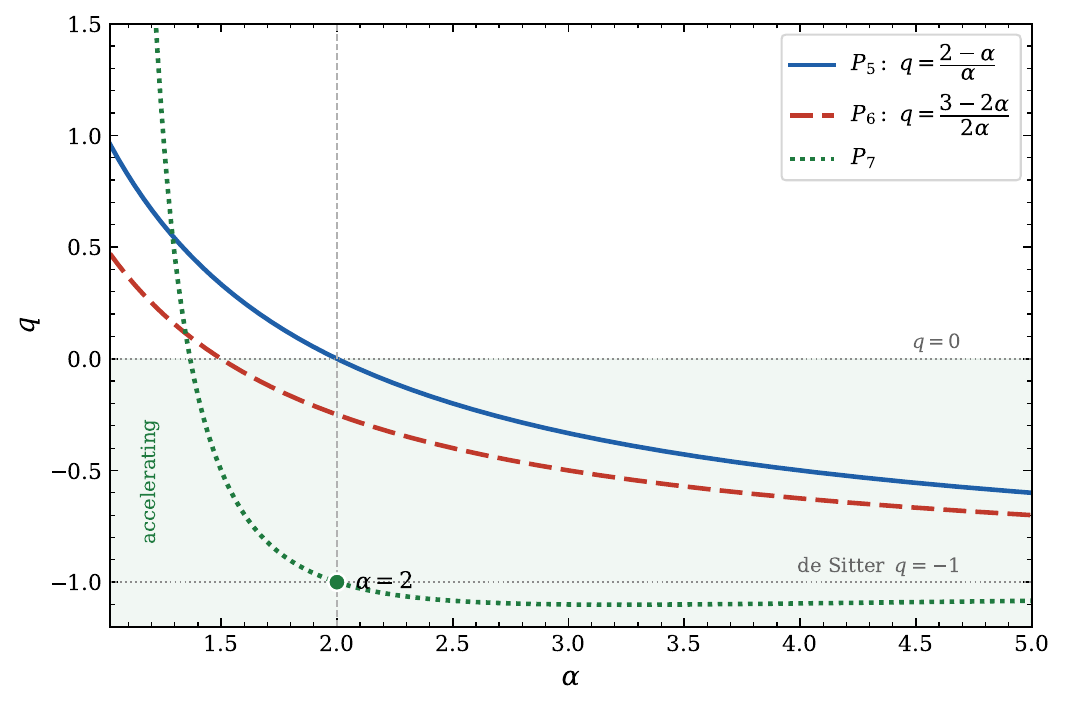}
\caption{Dependence of the deceleration parameter $q$ on the model parameter
$\alpha$ for the branches $P_5$,
$P_6$, and $P_7$. The shaded region marks the accelerating regime $q<0$, while
the dotted guides indicate $q=0$ and the de Sitter value $q=-1$. The branch
$P_7$ reaches $q=-1$ at $\alpha=2$, where it becomes degenerate with the de
Sitter configuration in the projected phase space.}
\label{fig:param_diag}
\end{figure}

Figure~\ref{fig:param_diag} summarizes how the parameter $\alpha$ controls the
kinematics of the scaling and accelerating branches. In particular, $P_7$ reaches
the de Sitter value at $\alpha=2$, motivating the use of this value as an
illustrative benchmark for the power-law phase-space analysis. This benchmark is
not the same model slice as the observational branch: the latter is the regular
Einstein--Hilbert branch $R-2\Lambda+\beta\chi$, whose background expansion has a
well-defined $\Lambda$CDM limit as $\beta\to0$.

\paragraph{Summary.}
The reduced power-law model $f(R,\chi)=R^{\alpha}+\beta\chi$ exhibits a rich
phase-space structure with radiation-like boundary configurations, intermediate
scaling saddles in restricted admissibility windows, and accelerating de
Sitter-like endpoints. For the illustrative benchmark $\alpha=2$, the two
accelerating branches $P_4$ and $P_7$ merge into a single de Sitter point in the
region $R\neq0$, with reduced-Jacobian spectrum $\{-4,-3,-3,0\}$; the vanishing
eigenvalue makes this point non-hyperbolic at exactly $\alpha=2$. Away from the
benchmark the branches separate, and $P_7$ becomes a genuine stable de Sitter
node (all four eigenvalues negative in the admissible range) while the
stability of $P_4$ is range-dependent, typically a saddle but not for all
$\alpha$ (e.g., $\alpha<0$). Several early-time and scaling points should therefore be interpreted as boundary or formally existing configurations rather than as a complete regular radiation-to-matter-to-de Sitter sequence. The phase-space analysis provides qualitative motivation for considering Ricci-tensor-squared corrections, while the direct observational constraints are obtained separately for the regular branch $R-2\Lambda+\beta\chi$.

\section{Observational Constraints}\label{sec:obs}

Having analysed the phase-space structure of the power-law family
$f(R,\chi)=R^{\alpha}+\beta\chi$ in the regular autonomous chart
$\alpha\neq1$, we now confront a regular low-curvature branch of the theory with
background cosmological data. The observational analysis is performed for the
Einstein--Hilbert branch supplemented by a cosmological constant and by the
Ricci-tensor-squared invariant,
\begin{equation}\label{eq:fobs}
f_{\rm obs}(R,\chi)=R-2\Lambda+\beta\chi .
\end{equation}
This branch corresponds to the $\alpha=1$ Einstein--Hilbert limit, augmented by
$\Lambda$ and by the Nash-type quadratic correction. It is not analysed with the
autonomous variables of Section~\ref{sec:dynamical}, because those variables
contain factors proportional to $(\alpha-1)^{-1}$ and become singular at
$\alpha=1$. The background Friedmann equations themselves, however, are regular
in this limit, since
\begin{equation}
f_R=1,\qquad f_\chi=\beta .
\end{equation}
We therefore solve the background equations directly in redshift space for the
branch \eqref{eq:fobs}.

As in Eq.~\eqref{eq:dimensionless_f}, the expression \eqref{eq:fobs} is written
in curvature units set by the reference scale $R_\star$. Equivalently, in
dimensionful notation one may write
\begin{equation}
f_{\rm obs}(R,\chi)
=
R_\star\left[
\frac{R}{R_\star}
-2\frac{\Lambda}{R_\star}
+\beta\frac{\chi}{R_\star^2}
\right],
\end{equation}
and suppress the explicit factors of $R_\star$ afterwards. The parameter
$\beta$ used in the likelihood analysis is therefore dimensionless.

The inclusion of the cosmological constant is essential for the observational
branch. Without the $-2\Lambda$ term, the model $R+\beta\chi$ would reduce to GR
plus a small higher-curvature correction and would not reproduce the observed
late-time acceleration in the limit of small $\beta$. With the term
$-2\Lambda$, by contrast, the limit $\beta\to0$ is exactly the flat
$\Lambda$CDM background. Thus $\Lambda$CDM is not merely an external comparison
model in this section, but the nested limit of the observational branch. The
parameter $\beta$ quantifies the allowed background-level departure from this
limit induced by the invariant $\chi=R_{\mu\nu}R^{\mu\nu}$.

\subsection{Reduced background branch}\label{subsec:reduced_branch}

The full theory containing $R_{\mu\nu}R^{\mu\nu}$ is higher order at the level of
the field equations. In the present likelihood analysis we constrain the reduced
background branch selected by continuity with $\Lambda$CDM as $\beta\to0$. This
should be understood as an effective background-level prescription rather than a
complete treatment of all higher-derivative modes. A full perturbative analysis,
including the additional propagating degrees of freedom, is left for future work.

For the branch \eqref{eq:fobs}, the reduced background equation can be written as
a quadratic algebraic equation for $dE/dz$, where $E(z)=H(z)/H_0$. We use the root
continuously connected to the standard $\Lambda$CDM behaviour and write
\begin{equation}\label{eq:obs_ode}
\frac{dE}{dz}
=
\frac{-B-\mathrm{sgn}(\beta)\sqrt{B^2-4AC}}{2A},
\end{equation}
with
\begin{align}
A &= -6\beta(1+z)^2E^2,\\
B &= -48\beta(1+z)E^3,\\
C &= 3E^2-\lambda-3\Omega_m(1+z)^3+72\beta E^4 .
\end{align}
Here $\lambda\equiv\Lambda/H_0^2$. These coefficients follow from the $00$
component of the reduced field equations of the observational branch with
$f_R=1$ and $f_\chi=\beta$; every $\beta$-dependent term is strictly
proportional to $\beta$, so the equation reduces smoothly to $\Lambda$CDM as
$\beta\to0$. The physically relevant root is the one that remains continuously
connected to the standard $\Lambda$CDM slope
$dE/dz=\tfrac{3}{2}\Omega_m(1+z)^2/E$ as $\beta\to0$; in practice this selects
the $-\sqrt{\cdot}$ branch for $\beta>0$ and the $+\sqrt{\cdot}$ branch for
$\beta<0$, as encoded by the $\mathrm{sgn}(\beta)$ factor in
Eq.~\eqref{eq:obs_ode}. At $\beta=0$ the quadratic expression becomes formally
singular because $A=0$; the nested $\Lambda$CDM case is therefore evaluated
through its analytic $\beta\to0$ limit rather than through the quadratic formula
itself.

For each pair $(\Omega_m,\beta)$, the value of $\lambda$ is not sampled
independently. It is fixed by the normalization condition $E(0)=1$. Numerically,
we determine $\lambda$ by a shooting procedure: starting from the matching value
at $z_{\rm match}=10$, the reduced background equation is integrated backwards to
$z=0$, and $\lambda$ is adjusted with a bracketing root finder until the
normalization condition is satisfied.

The modified branch is used only over the late- and intermediate-redshift
interval
\begin{equation}
0\le z\le z_{\rm match},\qquad z_{\rm match}=10 .
\end{equation}
For $z>z_{\rm match}$, the distance calculation is matched to a standard
radiation+matter+$\Lambda$ background,
\begin{equation}\label{eq:egrrad}
E_{\rm GR+rad}^2(z)
=
\Omega_r(1+z)^4+\Omega_m(1+z)^3+1-\Omega_m-\Omega_r ,
\end{equation}
where
\begin{equation}\label{eq:omegar}
\Omega_r h^2
=
2.469\times10^{-5}\left(1+0.2271N_{\rm eff}\right),
\qquad
N_{\rm eff}=3.046 .
\end{equation}
This high-redshift matching is part of the background prescription used in the
likelihood. The Nash-type correction is constrained by low- and
intermediate-redshift geometry, while standard early-universe physics is retained
for the compressed CMB distance-prior likelihood.

\subsection{Numerical solution and precomputed grid}\label{subsec:grid}

Solving Eq.~\eqref{eq:obs_ode} at every step of the Markov chain would be
computationally expensive. We therefore adopt a precomputed-grid strategy. The
background solutions are computed on a regular grid with $80$ nodes in
\begin{equation}
\Omega_m\in[0.20,0.45],
\end{equation}
$60$ nodes in
\begin{equation}
\beta\in[-3\times10^{-4},5\times10^{-4}],
\end{equation}
and $2000$ redshift nodes over the modified-gravity interval
\begin{equation}
0\le z\le z_{\rm match}=10 .
\end{equation}
At each grid point, the shooting procedure fixes $\lambda$ through $E(0)=1$.

From these solutions we tabulate the expansion rate
$E_{\rm MG}(z;\Omega_m,\beta)$ and the dimensionless comoving-distance integral
\begin{equation}\label{eq:Ifunc}
\mathcal{I}_{\rm MG}(z;\Omega_m,\beta)
=
\int_0^z\frac{dz'}{E_{\rm MG}(z';\Omega_m,\beta)} .
\end{equation}
During MCMC sampling, $E_{\rm MG}(z)$ and $\mathcal{I}_{\rm MG}(z)$ are recovered
by linear regular-grid interpolation. Grid points for which the shooting
procedure fails, the integration becomes unstable, or the discriminant
$B^2-4AC$ in Eq.~\eqref{eq:obs_ode} becomes negative are rejected and assigned
zero likelihood.

We validated the precomputed grid against direct numerical integrations at
randomly selected off-grid points in the prior volume. Over the redshift range
used directly by SNe and BAO, $0\le z\le2.5$, the relative interpolation errors
in both $E(z)$ and $\mathcal{I}_{\rm MG}(z)$ remain below $0.1\%$. Halving the
grid spacing changes the posterior means by less than one tenth of the quoted
$1\sigma$ uncertainties.

For distances beyond the matching redshift, the integral is evaluated as
\begin{equation}\label{eq:Izstar_match}
\mathcal{I}(z)
=
\mathcal{I}_{\rm MG}(z_{\rm match})
+
\int_{z_{\rm match}}^{z}\frac{dz'}{E_{\rm GR+rad}(z')},
\qquad
z>z_{\rm match}.
\end{equation}
Thus the modified branch enters the CMB distance calculation only through the
low-redshift part of the line-of-sight distance. The high-redshift contribution
and the sound horizon are computed with the standard radiation-aware background.

\subsection{MCMC setup}\label{subsec:mcmc_setup}

We sample the posterior distribution with a Markov Chain Monte Carlo algorithm
using a Metropolis--Hastings sampler
\citep{10.1093/biomet/57.1.97,1953JChPh..21.1087M}, as implemented in the
\texttt{Cobaya} package \citep{Torrado_2021}. The sampled parameters are the
present Hubble constant $H_0$, the matter density $\Omega_m$, the physical baryon
density $\omega_bh^2=\Omega_bh^2$, and the quadratic-curvature deviation
parameter $\beta$. The dimensionless cosmological constant
$\lambda=\Lambda/H_0^2$ is determined by the shooting normalization and is not
sampled independently.

We assume flat priors
\begin{align}
&H_0\in[55,80],
\qquad
\Omega_m\in[0.20,0.45], \nonumber\\
&\omega_bh^2\in[0.018,0.026],
\qquad
\beta\in[-3\times10^{-4},5\times10^{-4}] .
\end{align}
Convergence is monitored with the Gelman--Rubin statistic
\citep{1992StaSc...7..457G,Brooks01121998}, and chains are considered converged
when $R-1<0.01$. Burn-in is determined \emph{a posteriori} from the integrated
autocorrelation time $\tau$ of each parameter, discarding the first
$\sim10\tau_{\max}$ samples. The final chains contain an effective sample size of
order $\mathcal{O}(3000)$ independent draws.

\subsection{SNe Ia}\label{subsec:sne}

For the low-redshift expansion history we employ the Pantheon+ compilation of
Type Ia supernovae \citep{Scolnic:2021amr,Brout:2022vxf}. We apply the standard
cut $z_{\rm HD}>0.01$ to remove very nearby objects most affected by peculiar
velocities and host-galaxy calibration, and apply the same mask to the full
statistical-plus-systematic covariance matrix. The resulting sample contains
$1590$ light-curve measurements. Retaining the off-diagonal terms of the
covariance matrix accounts for correlations among supernovae and avoids
underestimating the parameter uncertainties.

The theoretical distance modulus is
\begin{equation}\label{mu}
\mu_{\rm th}(z)
=
5\log_{10}\!\left[D_L(z)\right]+25,
\end{equation}
with the luminosity distance
\begin{equation}\label{dl}
D_L(z)
=
(1+z)\frac{c}{H_0}\mathcal{I}_{\rm MG}(z;\Omega_m,\beta),
\end{equation}
over the Pantheon+ redshift range. To eliminate the degeneracy between the
supernova absolute magnitude and $H_0$, we analytically marginalize over the
unknown intercept. This gives
\begin{equation}\label{snlikelihood}
-2\ln\mathcal{L}_{\rm SN}
=
\Delta\boldsymbol{\mu}^{T}C^{-1}\Delta\boldsymbol{\mu}
-
\frac{
\left(\mathbf{1}^{T}C^{-1}\Delta\boldsymbol{\mu}\right)^2
}
{\mathbf{1}^{T}C^{-1}\mathbf{1}}
+
\ln\!\left(
\frac{\mathbf{1}^{T}C^{-1}\mathbf{1}}{2\pi}
\right),
\end{equation}
where
$\Delta\boldsymbol{\mu}=\boldsymbol{\mu}_{\rm th}-\boldsymbol{\mu}_{\rm obs}$,
$C$ is the total covariance matrix, and $\mathbf{1}$ is a vector of ones. This
marginalization makes the supernova likelihood insensitive to the absolute
distance scale, so that the SNe sample constrains the shape of the expansion
history.

\subsection{BAO}\label{subsec:bao}

Baryon acoustic oscillations provide a standard ruler that anchors the expansion
history at intermediate redshifts. We use the consensus BOSS/eBOSS measurements
\citep{eBOSS:2020yzd,eBOSS:2019dcv} at the five effective redshifts
\begin{equation}
z=\{0.38,\,0.51,\,0.698,\,1.48,\,2.334\},
\end{equation}
which provide the transverse comoving distance and the radial Hubble distance in
units of the sound horizon at the drag epoch, $D_M/r_d$ and $D_H/r_d$. The model
predictions are
\begin{equation}\label{dm}
D_M(z)
=
\frac{c}{H_0}\mathcal{I}_{\rm MG}(z;\Omega_m,\beta),
\qquad
D_H(z)
=
\frac{c}{H_0E_{\rm MG}(z;\Omega_m,\beta)} .
\end{equation}
The drag-epoch sound horizon $r_d(\omega_m,\omega_b)$ is computed using the
standard fitting formula.

In the present implementation, we approximate the BAO covariance as diagonal,
using the quoted
\\one-dimensional uncertainties for the transverse and radial
measurements. The BAO likelihood is therefore
\begin{align}\label{baolikelihood}
-2\ln\mathcal{L}_{\rm BAO}
=\ &
\sum_i
\left[
\frac{
(D_M/r_d)_{{\rm th},i}-(D_M/r_d)_{{\rm obs},i}
}
{\sigma_{D_M/r_d,i}}
\right]^2 \nonumber\\
&+
\sum_i
\left[
\frac{
(D_H/r_d)_{{\rm th},i}-(D_H/r_d)_{{\rm obs},i}
}
{\sigma_{D_H/r_d,i}}
\right]^2 .
\end{align}
This diagonal treatment matches the numerical likelihood used in the present
analysis. Given that the BAO data enter here through a limited set of consensus
distance measurements, this approximation is not expected to alter the
qualitative background-level conclusions. A future refinement should incorporate
the full consensus covariance matrix, including the transverse--radial
correlations.

\subsection{CMB distance priors}\label{subsec:cmb}

To include early-universe geometric information without running a full Boltzmann
solver, we adopt the compressed CMB likelihood based on the Planck~2018 distance
priors \citep{Planck:2018vyg,Chen:2018dbv}. These summarize the acoustic peak
geometry through the shift parameter $R$, the acoustic scale $\ell_A$, and the
physical baryon density $\omega_bh^2$. We use
\begin{equation}\label{cmb_mean}
\mathbf{v}_{\rm Planck}
=
(R,\ell_A,\omega_bh^2)
=
(1.74963,301.80845,0.02237),
\end{equation}
together with the corresponding covariance matrix.

The theoretical quantities are
\begin{equation}\label{cmbvec}
R
=
\sqrt{\Omega_m}\,\mathcal{I}(z_\ast),
\qquad
\ell_A
=
\pi\,\frac{c}{H_0}\frac{\mathcal{I}(z_\ast)}{r_s(z_\ast)} ,
\end{equation}
where $z_\ast$ is the redshift of photon decoupling computed from the
Hu--Sugiyama fitting formula. We do not extrapolate the modified-gravity solution
to recombination. Instead, the distance to last scattering is computed using
\begin{equation}\label{eq:cmb_distance_matching}
\mathcal{I}(z_\ast)
=
\mathcal{I}_{\rm MG}(z_{\rm match})
+
\int_{z_{\rm match}}^{z_\ast}
\frac{dz}{E_{\rm GR+rad}(z)} .
\end{equation}
The sound horizon is evaluated with the same standard high-redshift background,
\begin{equation}\label{eq:sound_horizon}
r_s(z_\ast)
=
\frac{c}{H_0}
\int_{z_\ast}^{z_{\rm max}}
\frac{c_s(z)/c}{E_{\rm GR+rad}(z)}\,dz,
\qquad
z_{\rm max}=2\times10^5,
\end{equation}
where $c_s(z)$ is the standard baryon--photon sound speed. The CMB contribution
to the likelihood is
\begin{align}\label{cmblikelihood}
-2\ln\mathcal{L}_{\rm CMB}
=\ &
\left(\mathbf{v}-\mathbf{v}_{\rm Planck}\right)^T
\mathsf{C}_{\rm CMB}^{-1}
\left(\mathbf{v}-\mathbf{v}_{\rm Planck}\right), \nonumber\\
&\mathbf{v}=(R,\ell_A,\omega_bh^2).
\end{align}

The use of compressed CMB distance priors in a modified-gravity context should be
viewed as an approximation. In the present analysis, the Nash-type correction is
constrained as a late-time background modification, while the pre-recombination
expansion, decoupling physics, and sound horizon are kept standard through the
matching prescription above. Thus the CMB likelihood mainly fixes the
high-redshift normalization of the distance scale and the comoving distance to
last scattering. A complete treatment of the CMB anisotropies in this theory
would require implementing the perturbation equations in a Boltzmann solver and
is beyond the scope of the present background-level analysis.

\subsection{Combined likelihood}\label{subsec:combined_likelihood}

The three probes are statistically independent to an excellent approximation, so
the total log-likelihood is taken as the sum of the individual contributions,
\begin{equation}\label{alllikelihood}
\ln\mathcal{L}_{\rm total}
=
\ln\mathcal{L}_{\rm SN}
+
\ln\mathcal{L}_{\rm BAO}
+
\ln\mathcal{L}_{\rm CMB}.
\end{equation}
The joint dataset combines low-redshift supernova distances,
intermediate-redshift BAO measurements, and the compressed CMB distance scale.
It therefore provides a strong background-level constraint on the deviation
parameter $\beta$, which controls the departure of the observational branch
\eqref{eq:fobs} from the nested $\Lambda$CDM limit.

\section{Results and Comparison with Observations}\label{sec:results}

\begin{figure*}[htbp]
\centering
\includegraphics[width=0.55\textwidth]{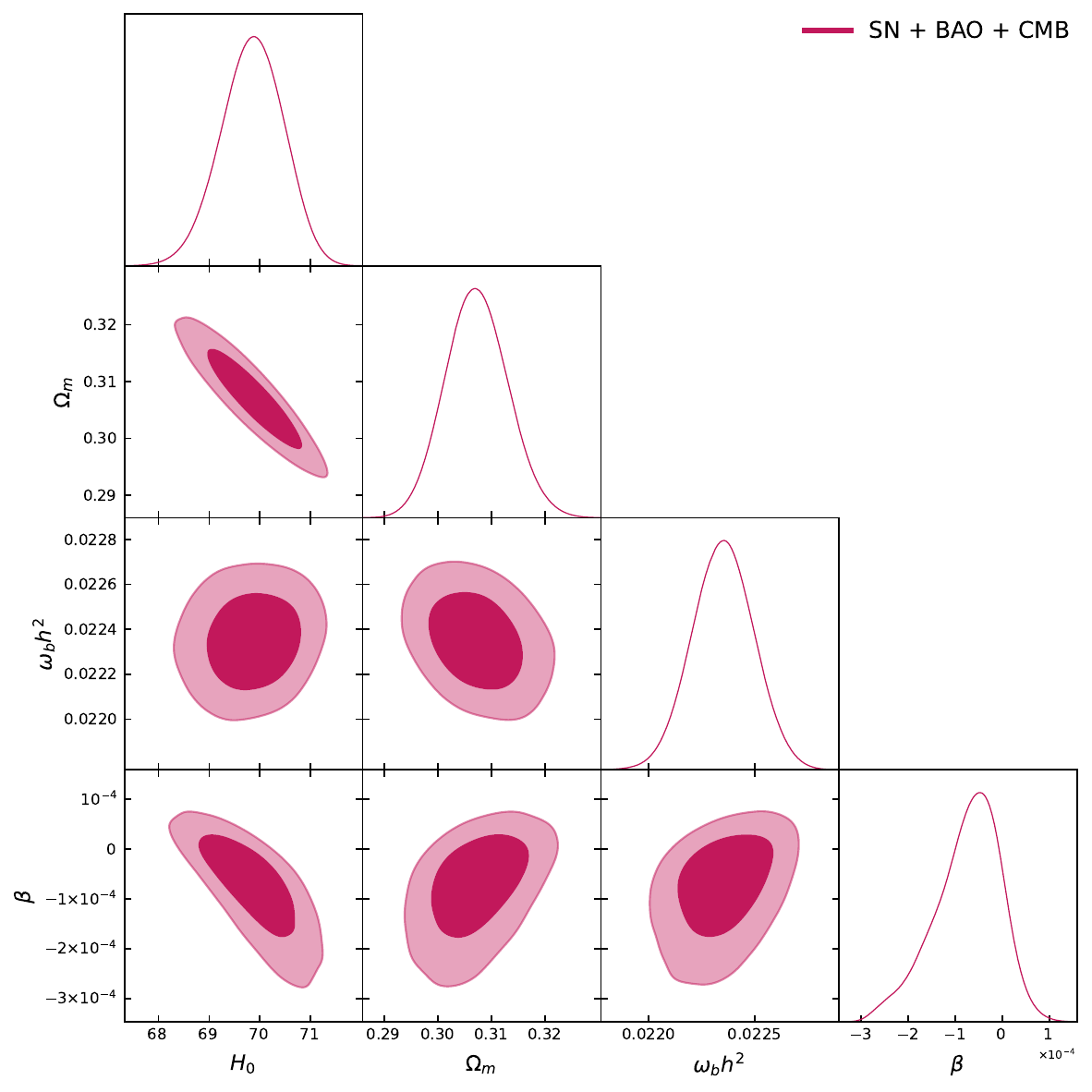}
\caption{Marginalized one- and two-dimensional posterior distributions with
$1\sigma$ and $2\sigma$ contours for the parameters
$\{H_0,\Omega_m,\omega_b h^2,\beta\}$ of the observational branch
$R-2\Lambda+\beta\chi$, obtained from the joint SNe+BAO+CMB chain. The posteriors
are well localized once the three background probes are combined.}
\label{fig:corner_frx}
\end{figure*}

We now present the cosmological constraints obtained from the joint
SNe\,(Pantheon+)\,+\,BAO\,+\,CMB analysis of the observational branch
\[
f_{\rm obs}(R,\chi)=R-2\Lambda+\beta\chi .
\]
The marginalized parameter constraints are collected in
Table~\ref{tab:bestfit_frx}, and the corresponding joint posterior distribution
is displayed in Figure~\ref{fig:corner_frx}. The corner plot was produced with
the \texttt{triangle\_plot} routine of the \texttt{GetDist} package
\citep{lewis2019getdistpythonpackageanalysing}.

\begin{table}[htbp]
\centering
\renewcommand{\arraystretch}{1.25}
\begin{tabular}{lccc}
\toprule
Parameter & Best fit & Mean & $68\%$ C.L. \\
\midrule
$H_0\ [\mathrm{km\,s^{-1}\,Mpc^{-1}}]$ & $69.710$ & $69.860$ & $\pm0.62$ \\
$\Omega_m$            & $0.3082$  & $0.3071$  & $\pm0.0058$ \\
$\omega_b h^2$        & $0.02236$ & $0.02235$ & $\pm0.00014$ \\
$\beta\ [10^{-5}]$    & $-4.5$    & $-6.6$    & $^{+6.0}_{-8.1}$ \\
\bottomrule
\end{tabular}
\caption{Best-fit values, posterior means, and $68\%$ confidence intervals for
the observational branch $f_{\rm obs}(R,\chi)=R-2\Lambda+\beta\chi$ from the
joint SNe+BAO+CMB analysis. The deviation parameter is quoted in units of
$10^{-5}$ and measures the departure from the nested $\Lambda$CDM limit
$\beta=0$.}
\label{tab:bestfit_frx}
\end{table}

The inferred Hubble constant,
\[
H_0=69.86\pm0.62\ \mathrm{km\,s^{-1}\,Mpc^{-1}},
\]
and matter density,
\[
\Omega_m=0.307\pm0.006,
\]
are close to the Planck~2018 $\Lambda$CDM values \citep{Planck:2018vyg}: the
matter density agrees with Planck to within about $1\sigma$, while $H_0$ lies
somewhat higher (see below). These values largely
reflect the strong constraining power of the CMB distance prior and BAO
standard-ruler information in fixing the high-redshift normalization of the
background expansion history. The physical baryon density is recovered with
sub-percent precision,
\[
\omega_bh^2=0.02235\pm0.00014,
\]
consistent with the value preferred by the acoustic-scale information entering
the compressed CMB likelihood.

The inferred value of $H_0$ sits slightly above the Planck-inferred value but
well below local distance-ladder determinations. Therefore, within the reduced
background-level prescription adopted here, the Nash-type quadratic correction
does not by itself resolve the Hubble tension. Its main effect is instead a small
late-time deformation of the expansion history around the nested
$\Lambda$CDM limit.

\begin{figure*}[htbp]
\centering
\includegraphics[width=0.5\textwidth]{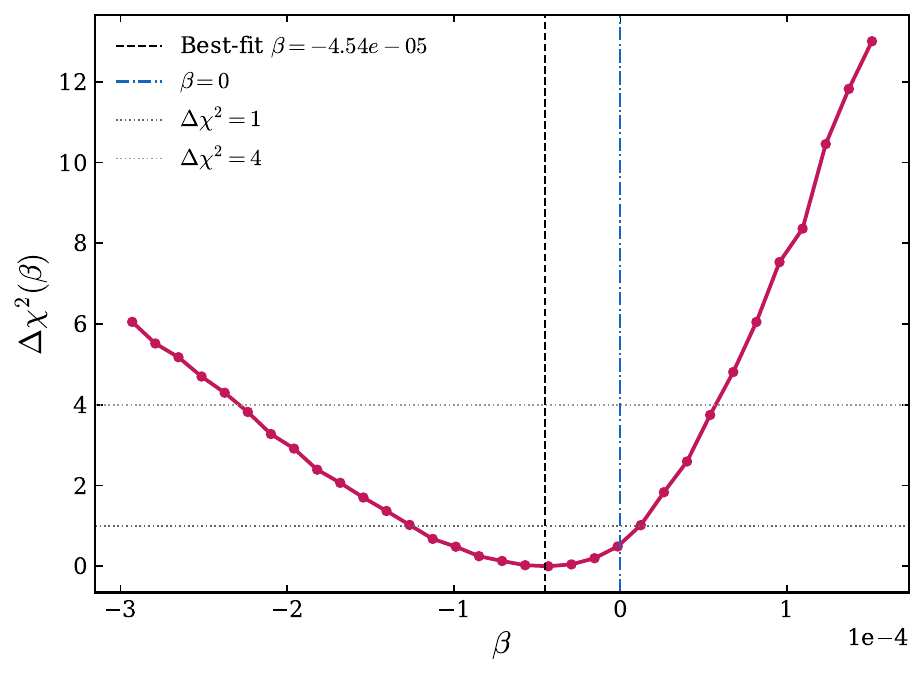}
\caption{Profile $\Delta\chi^2(\beta)$ reconstructed from the MCMC chain by
minimizing $\chi^2$ within bins of $\beta$ for the observational branch
$R-2\Lambda+\beta\chi$. The dashed vertical line marks the best-fit value, the
dot-dashed line indicates the nested $\Lambda$CDM limit $\beta=0$, and the dotted
horizontal lines show the $\Delta\chi^2=1$ and $\Delta\chi^2=4$ thresholds.}
\label{fig:profile_beta}
\end{figure*}

The central observational result concerns the deviation parameter $\beta$. We
obtain
\begin{equation}\label{eq:betaresult}
\beta=\left(-6.6^{+6.0}_{-8.1}\right)\times10^{-5}
\qquad (68\%\ \mathrm{C.L.}) .
\end{equation}
The posterior peaks at a small negative value of $\beta$. Although the marginalized $68\%$ credible interval (from $-14.7\times10^{-5}$ to $-0.6\times10^{-5}$) strictly excludes zero, the profile likelihood in Figure~\ref{fig:profile_beta} demonstrates that $\Delta\chi^2(\beta=0) < 1$. Therefore, from a profile-likelihood perspective, the nested $\Lambda$CDM limit $\beta=0$ remains consistent with the data at the $1\sigma$ level. The data therefore show no statistically significant preference for a nonzero Ricci-tensor-squared correction: the quadratic deviation is tightly constrained around the standard-model value $\beta=0$. This constraint
should itself be interpreted cautiously, since the likelihood constrains a
reduced $\Lambda$CDM-connected background branch, uses compressed CMB distance
priors, and treats the BAO covariance diagonally. Thus Eq.~\eqref{eq:betaresult}
is best understood as a background-level upper bound on the size of the
correction rather than as evidence for a departure from $\Lambda$CDM.

The posterior support remains close to the standard cosmological model. In
particular, the allowed values satisfy
\[
|\beta|\lesssim 3\times10^{-4}
\]
across the prior region favoured by the data. Current geometric probes therefore
confine the Nash-type correction to a narrow neighbourhood of $\Lambda$CDM.

\subsection{Goodness of fit and model selection}\label{subsec:model_selection}

At the best-fit point of the Nash-type branch, the total chi-square is
\[
\chi^2_{\min}=1423.43 .
\]
The dataset contains
\[
N_{\rm data}=1603
\]
points, consisting of $1590$ Pantheon+ supernovae, $10$ BAO distance
measurements, and $3$ CMB distance-prior quantities. The Nash-type observational
branch has $k=4$ sampled parameters,
\[
\{H_0,\Omega_m,\omega_bh^2,\beta\},
\]
whereas the nested $\Lambda$CDM limit corresponds to fixing $\beta=0$ and has
$k=3$ sampled parameters.

We compare the two models using the Akaike and Bayesian information criteria,
\begin{equation}\label{eq:aicbic}
\mathrm{AIC}=\chi^2_{\min}+2k,
\qquad
\mathrm{BIC}=\chi^2_{\min}+k\ln N_{\rm data}.
\end{equation}
The resulting comparison is summarized in Table~\ref{tab:model_comparison}.

\begin{table}[htbp]
\centering
\renewcommand{\arraystretch}{1.25}
\begin{tabular}{lccccc}
\toprule
Model & $k$ & $\chi^2_{\rm tot}$ & $\Delta\chi^2$ & AIC & BIC \\
\midrule
$\Lambda$CDM $(\beta=0)$ & $3$ & $1423.98$ & $0$     & $1429.98$ & $1446.12$ \\
$R-2\Lambda+\beta\chi$   & $4$ & $1423.43$ & $-0.55$ & $1431.43$ & $1452.95$ \\
\bottomrule
\end{tabular}
\caption{Model comparison between the nested $\Lambda$CDM limit and the
Nash-type observational branch. The one-parameter extension improves the minimum
chi-square by only $\Delta\chi^2\simeq-0.55$ relative to the fixed $\beta=0$
case; the $\beta=0$ chi-square is obtained from the profile
$\Delta\chi^2(\beta)$ of the chain (Figure~\ref{fig:profile_beta}) evaluated at
$\beta=0$. Both the AIC and the BIC therefore favour $\Lambda$CDM once the
additional parameter is penalized.}
\label{tab:model_comparison}
\end{table}

The Nash-type branch has a reduced chi-square
\[
\frac{\chi^2_{\min}}{N_{\rm data}-k}\simeq0.89,
\]
which indicates an acceptable background-level fit to the combined dataset. The
improvement
\[
\Delta\chi^2\simeq -0.55
\]
relative to the nested $\Lambda$CDM limit is marginal and is achieved at the cost
of one additional parameter. This gives
\[
\Delta\mathrm{AIC}\simeq +1.5,
\qquad
\Delta\mathrm{BIC}\simeq +6.8,
\]
so both information criteria favour the nested $\Lambda$CDM limit over the
extended branch. We consequently interpret the model-selection results as
showing no support for the one-parameter extension: the additional freedom
provided by $\beta$ does not yield a statistically meaningful improvement of the
fit, and the data are fully consistent with $\Lambda$CDM at the background
level.

\subsection{Reconstructed expansion history}\label{subsec:ez_results}

\begin{figure*}[htbp]
\centering
\includegraphics[width=0.55\textwidth]{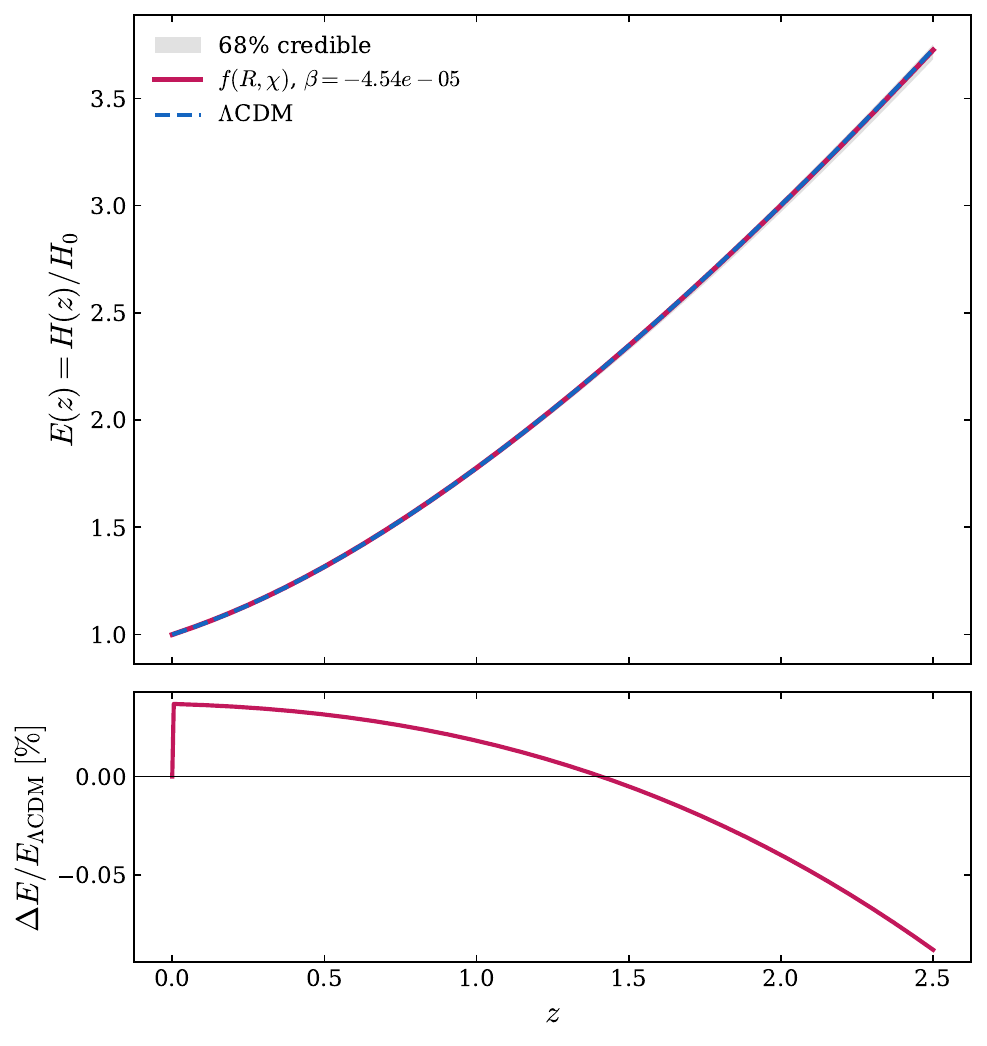}
\caption{Top: reconstructed dimensionless Hubble rate $E(z)=H(z)/H_0$ for the
best-fit observational branch $R-2\Lambda+\beta\chi$ (solid), with the shaded
band denoting the $68\%$ credible region, compared to the $\Lambda$CDM reference
(dashed). Bottom: fractional difference relative to $\Lambda$CDM. The deviation
remains below the percent level over the plotted redshift range.}
\label{fig:Ez_frx}
\end{figure*}

To visualize the impact of the inferred deviation parameter, Figure~\ref{fig:Ez_frx}
shows the reconstructed dimensionless Hubble rate $E(z)=H(z)/H_0$ for the
best-fit observational branch, together with its $68\%$ credible band. The result
is compared with the $\Lambda$CDM reference sharing the same matter density.

The lower panel of Figure~\ref{fig:Ez_frx} displays the fractional difference
relative to $\Lambda$CDM,
\[
\frac{\Delta E}{E_{\Lambda{\rm CDM}}}
=
\frac{E_{\rm Nash}(z)-E_{\Lambda{\rm CDM}}(z)}
{E_{\Lambda{\rm CDM}}(z)} .
\]
The deviation remains at the sub-percent level over the redshift range
$0\le z\le2.5$, with the largest difference occurring at low redshift where the
quadratic correction is most relevant. This confirms that the observational
branch remains very close to the standard expansion history while allowing a
small late-time deformation favoured by the combined background data.

\subsection{Evolution of the matter density parameter}\label{subsec:omz_results}

\begin{figure*}[htbp]
\centering
\includegraphics[width=0.55\textwidth]{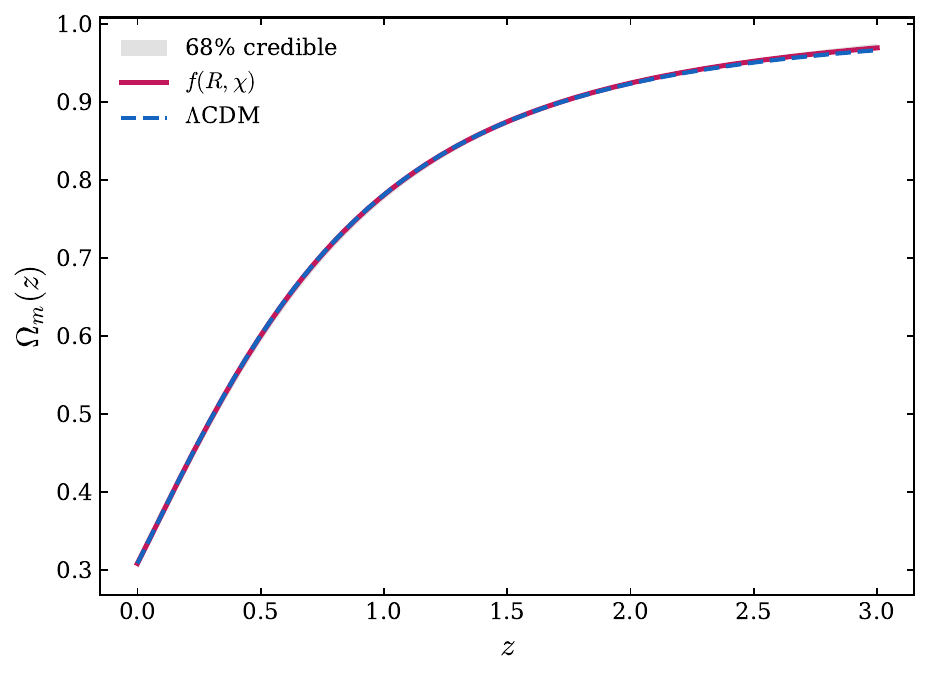}
\caption{Evolution of the matter density parameter $\Omega_m(z)$ for the
best-fit observational branch $R-2\Lambda+\beta\chi$ (solid) with its $68\%$
credible band, compared to the $\Lambda$CDM reference (dashed). The two
predictions track each other closely and approach the matter-dominated behaviour
at high redshift.}
\label{fig:Omz_frx}
\end{figure*}

The matter-sector evolution is shown in Figure~\ref{fig:Omz_frx}. We reconstruct
the effective matter density parameter as
\[
\Omega_m(z)=\frac{\Omega_m(1+z)^3}{E^2(z)} ,
\]
using the best-fit background and the posterior samples. At high redshift, the
model approaches the matter-dominated regime, while at low redshift
$\Omega_m(z)$ decreases as the effective dark sector becomes dominant.

The reconstructed trajectory remains very close to the corresponding
$\Lambda$CDM prediction throughout the plotted range. The small negative value of
$\beta$ produces only a sub-percent shift in the matter--dark-energy transition,
consistent with the tight constraint on the quadratic-curvature correction.

\subsection{Deceleration and effective equation of state}\label{subsec:qw_results}

The late-time acceleration is characterized by the deceleration parameter
\begin{equation}\label{eq:qz}
q(z)
=
-1+(1+z)\frac{1}{E}\frac{dE}{dz}.
\end{equation}
The corresponding effective total equation of state is
\begin{equation}
w_{\rm tot}(z)=\frac{2q(z)-1}{3}.
\end{equation}
We also reconstruct an effective dark-sector equation of state by defining
\begin{equation}
\rho_{\rm DE}^{\rm eff}(z)
=
E^2(z)-\Omega_m(1+z)^3 .
\end{equation}
The resulting $w_{\rm DE}(z)$ should be understood as a phenomenological
background reconstruction of the effective dark sector, not as the equation of
state of a fundamental fluid.

To obtain stable numerical derivatives, the functions entering $q(z)$ and
$w_{\rm DE}(z)$ are computed from a cubic-spline reconstruction of $E(z)$ on the
precomputed background grid rather than from unsmoothed finite differences. We
verified that refining the redshift grid changes the reconstructed $q(z)$ and
$w_{\rm DE}(z)$ curves by less than $0.5\%$ over the plotted range.

\begin{figure*}[htbp]
\centering
\includegraphics[width=0.55\textwidth]{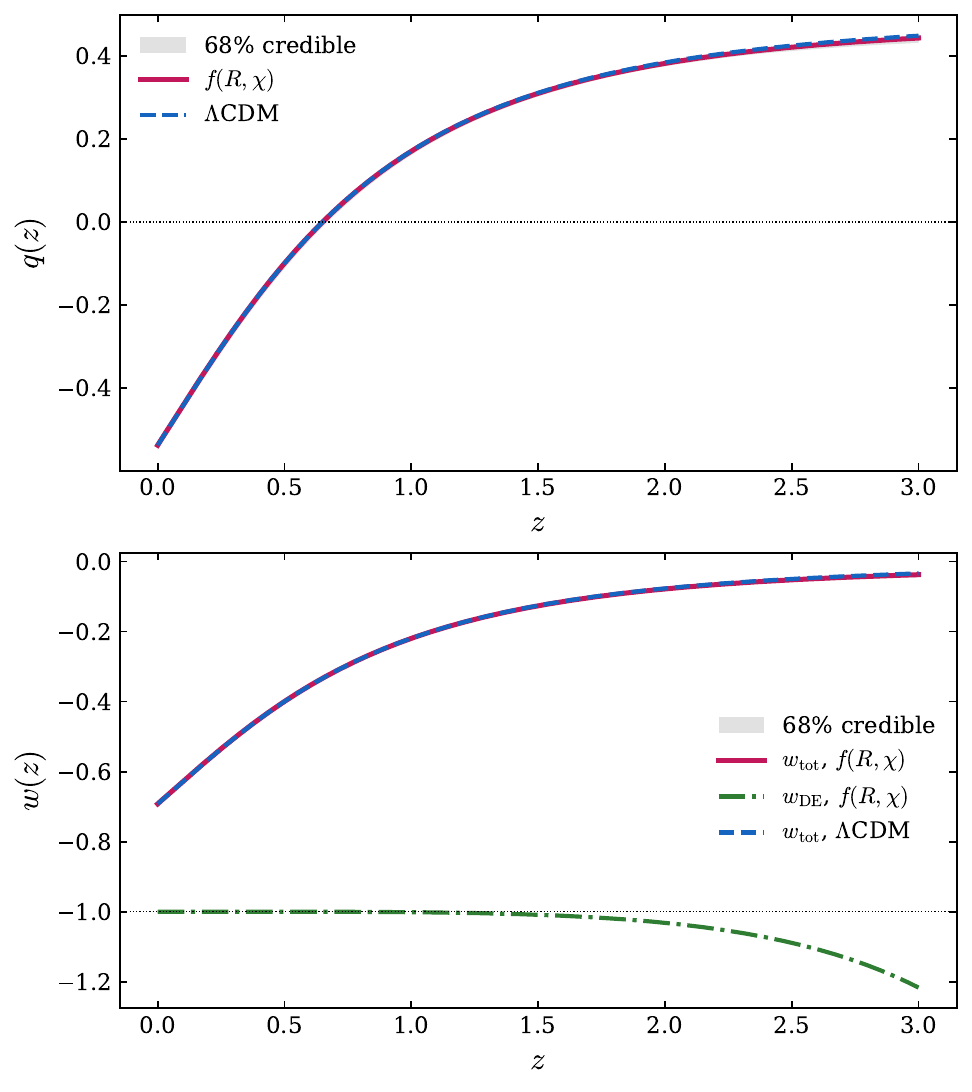}
\caption{Top: deceleration parameter $q(z)$ for the best-fit observational branch
$R-2\Lambda+\beta\chi$ (solid) and the $\Lambda$CDM reference (dashed), with the
$68\%$ credible band. Bottom: effective equation of state, showing the total
fluid $w_{\rm tot}(z)$ and the reconstructed effective dark-sector
$w_{\rm DE}(z)$ against the $\Lambda$CDM reference. Both quantities are obtained
from a cubic-spline reconstruction of $E(z)$.}
\label{fig:qw_frx}
\end{figure*}

The results are shown in Figure~\ref{fig:qw_frx}. The best-fit model undergoes
the deceleration-to-acceleration transition at
\[
z_t\simeq0.7,
\]
in close agreement with $\Lambda$CDM. The effective dark-sector equation of state
remains close to
\[
w_{\rm DE}=-1
\]
over the observed redshift range. The small negative best-fit value of $\beta$
therefore appears only as a mild low-redshift deviation from the
cosmological-constant behaviour, well within the $68\%$ credible band that
contains the $\Lambda$CDM limit $\beta=0$ and too small to constitute a
statistically significant background-level signature.

\subsection{Fit to the Pantheon+ Hubble diagram}\label{subsec:pantheon_fit}

\begin{figure*}[htbp]
\centering
\includegraphics[width=0.55\textwidth]{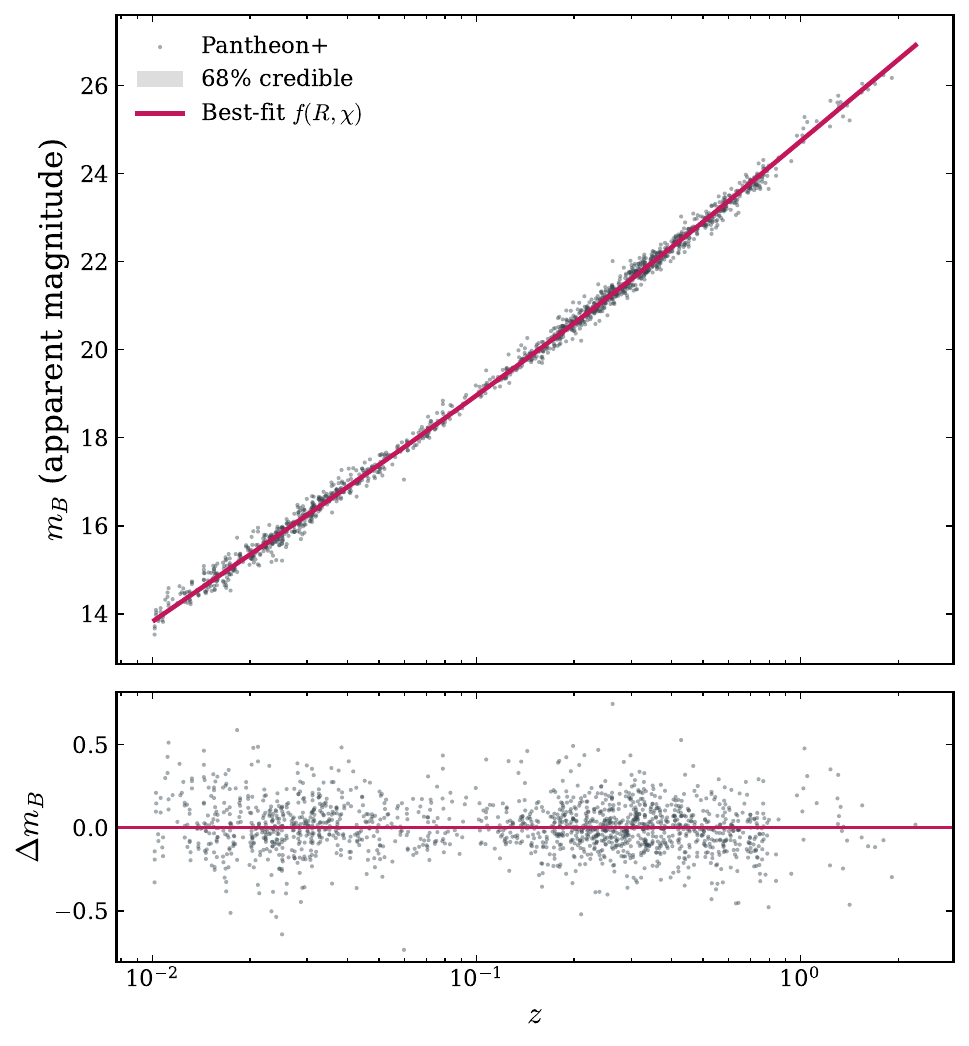}
\caption{Top: Pantheon+ apparent magnitudes and the best-fit prediction of the
observational branch $R-2\Lambda+\beta\chi$ (solid) with its $68\%$ credible
band, shown on a logarithmic redshift axis. Bottom: magnitude residuals
$\Delta m_B$ about the best-fit model.}
\label{fig:pantheon_frx}
\end{figure*}

Figure~\ref{fig:pantheon_frx} shows the Pantheon+ Hubble diagram with the
best-fit luminosity-distance relation overlaid, together with the magnitude
residuals. The observational branch reproduces the supernova data over the full
redshift range. The residuals are symmetrically distributed around zero and show
no visible systematic trend. This confirms that the small curvature correction
inferred from the combined likelihood is compatible with the low-redshift
supernova distance data at the background level.

In summary, the joint SNe+BAO+CMB analysis constrains the observational branch
$R-2\Lambda+\beta\chi$ to lie very close to the nested $\Lambda$CDM limit. For
the converged chains used in this analysis, the quadratic-curvature deviation
parameter is
\[
\beta=\left(-6.6^{+6.0}_{-8.1}\right)\times10^{-5},
\]
consistent with the nested $\Lambda$CDM limit $\beta=0$ at about the $1\sigma$
level and corresponding to a tight background-level upper bound on a small
Ricci-tensor-squared correction rather than to a detection. The remaining
cosmological parameters stay close to their standard values, and the
reconstructed expansion history, matter evolution, deceleration parameter, and
supernova fit all remain phenomenologically close to $\Lambda$CDM. Whether any
small curvature signature of this type is present at all must be tested with the
full BAO covariance, updated survey data, and a complete perturbation-level
analysis.

\section{Discussion and Conclusions}\label{sec:conclusions}

In this work we have investigated cosmological aspects of generalized Nash-type
gravity models involving the quadratic Ricci invariant
$\chi=R_{\mu\nu}R^{\mu\nu}$. The analysis was deliberately divided into two
complementary parts. First, we studied the phase-space structure of the power-law
family $f(R,\chi)=R^{\alpha}+\beta\chi$ in the regular autonomous-system chart
$\alpha\neq1$. Second, we confronted current background cosmological data with
the regular Einstein--Hilbert observational branch
$f_{\rm obs}(R,\chi)=R-2\Lambda+\beta\chi$, for which the background equations are
well defined and the limit $\beta\to0$ is exactly flat $\Lambda$CDM. The two
branches are therefore not treated as identical parameter slices: the power-law
sector is used to diagnose the qualitative dynamical role of the
Ricci-tensor-squared correction, while the Einstein--Hilbert branch is used for a
controlled background-level comparison with data around the nested $\Lambda$CDM
limit.

On the theoretical side, recasting the field equations of the power-law family as
a reduced four-dimensional autonomous system revealed a rich critical-point
structure. The flow contains radiation-like boundary configurations
($P_1,P_2,P_3,P_8$), intermediate scaling saddles ($P_5,P_6$) whose physical
admissibility is restricted to finite intervals of the exponent $\alpha$, and
accelerating branches ($P_4,P_7$) that organize the late-time sector. Because the
adopted autonomous variables contain factors proportional to $(\alpha-1)^{-1}$,
the Einstein--Hilbert case $\alpha=1$ cannot be treated in this chart. Moreover,
for the illustrative quadratic benchmark $\alpha=2$, the points located at
$x_2=0$ lie on the singular boundary $R=0$ of the chosen variables and should be
interpreted as limiting configurations rather than regular interior equilibria.
Within the regular region $R\neq0$, the late-time behaviour is governed by the
two accelerating branches $P_4$ and $P_7$. At the benchmark $\alpha=2$ they merge
into a single de Sitter point with reduced-Jacobian spectrum $\{-4,-3,-3,0\}$;
the vanishing eigenvalue makes this point non-hyperbolic, so linear theory alone
does not fix its stability along the degenerate direction. Away from $\alpha=2$
the branches separate, and the $P_7$ branch becomes a genuine stable de Sitter
node (all four eigenvalues negative), while $P_4$ turns into a saddle for most
values of $\alpha$, though its stability is range-dependent and negative real
parts occur in specific restricted intervals (e.g., $\alpha<0$). Thus the
power-law Nash-type sector naturally admits a stable accelerating endpoint, but
this result should not be read as a direct stability proof for the separate
observational branch.

On the observational side, we constrained the reduced background branch of
$f_{\rm obs}(R,\chi)=R-2\Lambda+\beta\chi$ using the joint
SNe\,(Pantheon+)\,+\,BAO\,+\,CMB distance-prior likelihood. The dimensionless
cosmological constant $\lambda=\Lambda/H_0^2$ was fixed at each grid point by the
normalization condition $E(0)=1$, so that $\beta$ is the only additional sampled
background parameter relative to flat $\Lambda$CDM. The modified branch was
integrated over $0\le z\le z_{\rm match}=10$ and then matched to a standard
radiation+matter+$\Lambda$ background for the high-redshift distance calculation.
Accordingly, the Planck~2018 CMB information was incorporated through compressed
distance priors rather than through a full modified-gravity Boltzmann analysis.
The BAO likelihood used in the numerical implementation was treated with diagonal
errors for the quoted transverse and radial measurements; incorporating the full
BAO covariance is a useful refinement for future work.

For the converged chains used in this analysis, the inferred background parameters remain close to the standard cosmological values, with $H_0=69.86\pm0.62\ \mathrm{km\,s^{-1}\,Mpc^{-1}}$, $\Omega_m=0.307\pm0.006$, and $\omega_bh^2=0.02235\pm0.00014$. The deviation parameter is constrained to
\begin{equation}
\beta=\left(-6.6^{+6.0}_{-8.1}\right)\times10^{-5},
\end{equation}
within the adopted reduced background prescription and prior range. The posterior therefore peaks at a small negative value of $\beta$. While the marginalized $68\%$ interval narrowly excludes zero, the profile likelihood reveals that $\beta=0$ falls within the $\Delta\chi^2 < 1$ region, indicating that the nested $\Lambda$CDM limit remains consistent with the data at the $1\sigma$ level. This should be interpreted as a tight background-level upper bound on a small Ricci-tensor-squared correction, not as evidence for modified gravity. Model-selection diagnostics reinforce this conclusion: the one-parameter
extension improves the minimum $\chi^2$ only marginally
($\Delta\chi^2\simeq-0.55$) relative to the fixed $\beta=0$ limit, and both the
AIC and the BIC favour the nested $\Lambda$CDM model once the extra parameter is
penalized.

The reconstructed expansion rate, matter-density evolution, deceleration
parameter, and effective dark-sector equation of state all show that the fitted
observational branch remains in the immediate neighbourhood of $\Lambda$CDM. The
transition to acceleration occurs at $z_t\simeq0.7$, and the effective dark-energy
equation of state stays close to $w_{\rm DE}=-1$ over the observed redshift range.
The inferred value of $H_0$ lies slightly above the Planck value but well below
the local distance-ladder determination, so the present Nash-type correction does
not by itself resolve the Hubble tension within the background-level setup
considered here.

Several limitations are central to the interpretation of these results. First,
the dynamical-system and observational analyses refer to different but related
branches of the generalized Nash-type framework. The de Sitter endpoint found in
the power-law sector (a stable node on the $P_7$ branch for $\alpha\neq2$, and
non-hyperbolic exactly at the benchmark $\alpha=2$ where $P_4$ and $P_7$
coincide) demonstrates the existence of stable accelerating endpoints in that
branch, but it does not establish the stability of the fitted Einstein--Hilbert
branch. Second, the observational likelihood
constrains a reduced $\Lambda$CDM-connected background branch and uses a
high-redshift matching prescription, compressed CMB distance priors, and a
diagonal BAO approximation. Third, a complete physical assessment of theories
containing $R_{\mu\nu}R^{\mu\nu}$ requires perturbation-level work, including the
analysis of ghost and gradient stability, gravitational-wave propagation, and
consistency with large-scale-structure growth. Future high-precision surveys such
as DESI, Euclid, and LSST, together with a full perturbative treatment and a
likelihood including the full BAO covariance, will be essential for tightening
the bound on $\beta$ and determining whether any small curvature signature of
this type is physically present or is merely absorbed into the standard-model
background.

%\appendix

%\section{Appendix information}

% -------------------------------------------------
% Bibliography
% -------------------------------------------------
\bibliographystyle{unsrtnat}
\bibliography{sample701}

\end{document}